\documentclass[journal]{IEEEtran}

\IEEEoverridecommandlockouts    
\usepackage{booktabs}

\usepackage{psfrag}
\usepackage{amsmath}
\usepackage{amssymb}

\usepackage{amsfonts}
\usepackage{latexsym}
\usepackage{graphicx}
\usepackage{amsmath,amssymb,amsfonts,bbm}

\usepackage{colortbl}
\usepackage{fancyhdr}
\usepackage{epstopdf}
\usepackage{float}

\usepackage{subcaption}
\usepackage[colorlinks=true,linkcolor=black,citecolor=black]{hyperref}
\DeclareMathOperator*{\argmin}{arg\,min}

\usepackage[usenames,dvipsnames,svgnames,table]{xcolor}

\def\tvdots{\raisebox{3pt}{$\scalebox{.75}{\vdots}$}}

\newtheorem{theorem}{Theorem}
\newtheorem{definition}{Definition}
\newtheorem{proposition}{Proposition}
\newtheorem{lemma}{Lemma}
\newtheorem{corollary}{Corollary}

\newtheorem{remark}{Remark}
\newtheorem{assumption}{Assumption}

\IEEEoverridecommandlockouts

\usepackage{soul}
\usepackage[ruled]{algorithm2e}

\newcommand{\R}{\mathbb{R}}
\newcommand{\Z}{\mathbb{Z}}
\newcommand{\N}{\mathbb{N}}

\newcommand{\mc}{\mathcal}

\newcommand{\Sw}{\boldsymbol{\mathcal{S}} }
\newcommand{\be}{\begin{equation}}
\newcommand{\ee}{\end{equation}}

\def\C{C}
\def\c{c}

\def\m{\boldsymbol{ \mathcal{A}}}
\def\mn{{\boldsymbol{\mathcal{A}}_{\nu}}}
\def\mnt{{\boldsymbol{\mathcal{A}}_{\nu_1,\nu_2}}}

\def\mnz{{\boldsymbol{\mathcal{A}}_{0,\nu}}}
\def\mnh{{\boldsymbol{\mathcal{A}}_{\frac{\nu}{2},\frac{\nu}{2}}}}

\def\Ie{\boldsymbol{\mathcal{I}}}

\def\Pn{P^{\nu}}
\def\Pen{{\boldsymbol{\mathcal{P}}_{\nu}}}
\def\Penh{{\boldsymbol{\mathcal{P}}_{\frac{\nu}{2}}}}
\def\ones{\frac{1}{N} \mathbbm{1}_{N}\mathbbm{1}_{N}^\top }
\def\one{ \mathbbm{1}_{N} }

\def\zl{z}

\def\ze{\boldsymbol{z}}
\def\xe{\boldsymbol{x}^\star}

\def\ae{\boldsymbol{r}}
\def\be{\boldsymbol{s}}
\def\a{r}
\def\b{s}

\graphicspath{{Figures/} }

\begin{document}
\title{ Network Aggregative Games and Distributed Mean Field Control via Consensus Theory}
\author{Francesca Parise, Sergio Grammatico, Basilio Gentile and  John Lygeros%
\thanks{The authors are with the Automatic Control Laboratory, ETH Zurich, Switzerland. 
E-mail addresses: \{\texttt{parisef}, \texttt{grammatico},  \texttt{gentileb}, \texttt{lygeros}\}\texttt{@control.ee.ethz.ch}.
}
}
\maketitle

\begin{abstract}
We consider network aggregative games to model and study multi-agent populations in which each rational agent is influenced by the aggregate behavior of its neighbors, as specified by an underlying network. Specifically, we examine systems where each agent minimizes a quadratic cost function, that depends on its own strategy and on a convex combination of the strategies of its neighbors, and is subject to personalized convex constraints. We analyze the best response dynamics and we propose  alternative distributed algorithms to  steer the strategies of the rational agents to a Nash equilibrium configuration. The convergence of these schemes is guaranteed under different sufficient conditions, depending on the matrices defining the cost and on the network. Additionally, we propose an extension to the network aggregative game setting that allows for multiple rounds of communications among the agents, and we illustrate how it can be combined with consensus theory to recover a solution to the mean field control problem in a distributed fashion, that is, without requiring the presence of a central coordinator. 
Finally, we apply our theoretical findings to study  a novel multi-dimensional, convex-constrained model of opinion dynamics and  a hierarchical demand-response scheme for  energy management in smart buildings, extending literature results.
\end{abstract}

\section{Introduction}
In recent years, there has been an increasing interest in modeling and control of populations of agents that interact through a network. 
If the agents are noncooperative and profit-maximizing, these systems can be studied combining ideas of game theory and network analysis. 
Traditionally, the literature on network game theory  has  focused on games in which each agent behavior is influenced by its \textit{one-to-one} interaction with the neighbors \cite{jackson2014games, galeotti2010network,szabo2007evolutionary,madeo2015game,comodistributed}. 
When the size of the population becomes very large, however, the analysis of these models may become computationally intractable. 
Moreover, in many applications involving large populations of agents
the well-being of an agent cannot be described as the overall result of individual two-player interactions, but should rather be modeled as a unique function depending on the agents strategy and on a quantity (either common to all agents or  agent-specific) that depends on the \textit{aggregate}  behavior of the entire population.

Examples of applications where the influence of the population behavior is the same for all agents are demand side management in smart grids~\cite{mohsenian-rad:10, bagagiolo:bauso:14, chen2014autonomous}, charging coordination of plug-in electric vehicles \cite{ma:callaway:hiskens:13, parise:colombino:grammatico:lygeros:14}, congestion control  \cite{barrera:garcia:15} and economic  markets  \cite{kizilkale:mannor:caines:12}. 
This type of interactions can be modeled using the theory of \textit{(generalized) aggregative games} since the payoff of each agent is  a function of its own strategy and of a common aggregator function, whose value depends on the strategies of all the players \cite{cornes2012fully}. In this framework, to compute its best response each agent needs to know the aggregate of all the other players' strategies; this may not be desirable when the population size is very large. One way to overcome this issue is to address large-size aggregative games by    assuming a continuum  of players that are influenced only by the population statistical distribution  \cite{huang:caines:malhame:07, lasry:lions:07}. These models can be efficiently analyzed by exploiting the so-called \textit{Mean Field} (MF) approximation. Specifically, several schemes have been proposed in the literature to coordinate the agents behavior in a decentralized fashion, by using a statistical description of the population \cite{huang:caines:malhame:07, bauso:pesenti:13}, or by means of a central coordinator \cite{ma:callaway:hiskens:13,grammatico:parise:colombino:lygeros:14}.

Classical aggregative and mean field games are based on the assumption that all the agents are affected by the population via the same aggregator function. 
On the other hand, we consider here situations where the interaction, even though not one-to-one as in typical network games, still possesses a well determined structure.
Examples where such structures appear are (generalized) quasi-aggregative games \cite{jensen:10} where each agent has a different interaction function,
  leader-follower games \cite{nourian:caines:malhame:huang:12} where some agents have more control authority or knowledge than others, games with local/agent-dependent cost functions \cite{huang:caines:malhame:10,grammatico2015cdc}, games where the agents are allowed to interact only locally \cite{chen2014autonomous,bauso2013distributed}, or games where the agents have different stubbornness \cite{stella2013opinion,bausoopinion}. 
Here, we consider   a setting in which each agent is influenced by the aggregate strategies of an agent-dependent subset of the population, which is defined by an underlying network. We refer to this type of games as \textit{Network Aggregative} (NA) \textit{games}. Among these, we restrict our attention to games in which each agent minimizes a quadratic cost function that depends on its own  strategy and on a convex combination of the strategies of its neighbors, that we refer to as the neighbors' aggregate state. Moreover, contrary to traditional (quasi-) aggregative games, 
we allow the aggregator function to be vector valued instead of scalar valued.

For this  class of systems, we are interested in  sets of strategies where no agent has interest in unilaterally deviating from its own behavior, that is, we aim at characterizing the Nash equilibria of the NA game. Moreover, we analyze the dynamics arising in a population where each agent updates synchronously its strategy in response to the strategies of its neighbors. The simplest type of dynamics that we investigate is the so-called Best Response (BR) dynamics, where the agents adopt at every iteration the strategy that minimizes their cost, given the current neighbors' aggregate state. 
Our first technical contribution is the derivation of conditions on the cost function and on the network structure that guarantee convergence of the BR dynamics to a Nash equilibrium. For cases where the BR dynamics are not guaranteed to converge, we propose different strategy update schemes with memory that can be implemented by the agents in a distributed  fashion and, under different conditions, ensure convergence to a Nash equilibrium.

Secondly, we show how the setting of NA games can be used to steer a population of agents to an almost Nash equilibrium of a deterministic  MF game by allowing multiple rounds of communications  between two  strategy updates. To this end, we use results from consensus theory to ensure that the agents can reconstruct the average of the strategies of the entire population by communicating with their neighbors. The advantage of this approach is that it does not require communications with all the agents in the population (as in aggregative games) making our method scalable as the population size increases, nor the presence of a central coordinator or a statistical description of the population (as in MF games).
The task of recovering a MF Nash equilibrium by allowing only local communications over a network has been already considered in the literature for the case of linear cost \cite{chen2014autonomous} or scalar state variables \cite{koshal2012gossip}, via  variational inequality approaches. Instead, we consider here a setting where each agent solves a multi-dimensional convex-constrained  optimization problem with quadratic cost function. 
A key aspect that differentiates our approach from more standard distributed optimization schemes is that in our setting, at every iteration of the algorithm, each agent selects its strategy as the BR to its current estimate of the population behavior. This is a fundamental feature for the applicability of our schemes in a population of non-cooperative agents. We also notice that the inverse problem of recovering consensus via MF games has been studied in~\cite{nourian2013nash,bauso2006non}.

The contributions of the paper are organized as follows. 
\begin{itemize}
\item As motivating application of NA games,  we formulate in Section \ref{sec:motivation} a novel extension of the classical Friedkin and Johnsen model of  opinion dynamics by allowing  each agent to update its opinions regarding a collection of different topics, linked together by convex and compact constraints. Moreover, we suggest a distributed implementation of a classic model of demand-response scheme for smart energy management, as motivating application for distributed MF control. 
\item In
Section~\ref{sec:problem_statement} we define NA games and illustrate their relations with aggregative and MF games. In Section~\ref{sec:problem_statement_3} we extend the definition of NA games by allowing multiple rounds of communications.
\item In Section \ref{sec:fixed_point} we introduce the concept of aggregation mapping, generalizing the ideas presented in \cite{grammatico:parise:colombino:lygeros:14}, and characterize the game-theoretical properties of the set of strategies computed as BR to its fixed points, for both  NA and MF games. 
\item In Section~\ref{sec:quest} we derive conditions guaranteeing the convergence of the BR dynamics and of  newly introduced update schemes to a fixed point of the aggregation mapping, generalizing the results in \cite{parise2015cdc}.   
\item 
In Section~\ref{sec:applications} we apply our theoretical findings to the applications introduced in Section  \ref{sec:motivation}, extending previous literature results.
\end{itemize}
Section~\ref{sec:conclusion} concludes the paper and highlights several possible extensions and applications.
Appendix~A presents some background definitions and results from operator theory;  Appendix~B contains all the proofs of the technical results.

\subsection*{Notation}
$\R$, $\R_{>0}$, $\R_{\geq 0}$ respectively denote the set of real, positive real, non-negative real numbers; $\N$ denotes the set of positive natural numbers; 
$2\N$ denotes the set of even numbers; 
$\Z$ denotes the set of integer numbers; for $a, b \in \Z$, $a \leq b$, $\Z[a,b] := [a,b] \cap \Z$. $I_n$ denotes the $n$-dimensional identity matrix; $\mathbbm{1}_{n}$ denotes the $n$-dimensional vector with all entries equal to $1$. For a given $Q \in \R^{n \times n}$, the positive definite notation  $Q \succ 0$ (as well as $ \succcurlyeq,\prec,\preccurlyeq$) implies $Q=Q^\top$.
For a given $Q \in \R^{n \times n}$, $Q \succ 0$, we denote by $\mathcal{H}_{Q}$ the Hilbert space $\R^n$ with inner product $\langle \cdot, \cdot \rangle_{Q}: \R^n \times \R^n \rightarrow \R$ defined as $\langle x, y \rangle_{Q} := x^\top Q y$, and  norm $\left\| \cdot \right\|_Q : \R^n \rightarrow \R_{\geq 0}$ defined as $\left\| x \right\|_Q := \sqrt{ x^\top Q x }$; $\|x\|:=\|x\|_{I_n}$. The projection operator in $\mathcal{H}_{Q}$, $\text{Proj}_{\mathcal{C}}^{Q} : \R^n \rightarrow \mathcal{C} \subseteq \R^n $, is defined as $\text{Proj}_{\mathcal{C}}^{Q}(x) := \argmin_{y \in \mathcal{C}} \left\|x-y \right\|_Q$.   $A \otimes B$ denotes the Kronecker product between matrices $A$ and $B$. Given a matrix $P$, we define $P_{ij}:=\left[P\right]_{ij}$ its element in position $(i,j)$ and for any $\nu\in\N$,  $P^\nu_{ij}$ denotes the element in position $(i,j)$ of the matrix $P^\nu$. The matrix $P$ is row-stochastic if $\sum_{j=1}^N P_{ij}=1$ for all $i\in\Z\left[1,N\right]$ and  doubly-stochastic if  $P$ and $P^\top$ are row-stochastic. The matrix $P$ is primitive if there exists $h\in\N$ such that $P^h$ is element-wise strictly positive. $\|P\|_Q:=\sup\left\{\frac{\|Px\|_{Q}}{\|x\|_{Q}}\mid x\in\R^n\setminus\{ 0\}\right\}$ denotes the induced matrix norm. We denote by $\Lambda(P)$ the set of eigenvalues of $P$,  by $\rho(P):=\max\left\{\ |\lambda| \mid \lambda\in\Lambda(P)\right\}$ its spectral radius and by $\sigma_{\text{max}}(P)$ its maximum singular value. If $P\succcurlyeq0$, we denote by $P^{\frac12}\succcurlyeq0$ its principal square root. Given $\mathcal{X} \subseteq \R^n$, $A \in \R^{n \times n}$ and $b\in\R^n$, $A \mathcal{X}+b$ denotes the set $\left\{ A x+b\in\R^n\mid \ x \in \mathcal{X} \right\}$; given $\mathcal{X}_1, \ldots, \mathcal{X}_N \subseteq \R^n$, $\mathcal{X}_{1\times N}:=\mathcal{X}_{1}\times \hdots\times \mathcal{X}_{N}$, $\textstyle \frac{1}{N}( \sum_{i=1}^{N} \mathcal{X}_i )  := \{  \frac{1}{N} \sum_{i=1}^{N} x_i \in \R^n \mid \ x_i \in \mathcal{X}_i, \ \forall i \in \Z[1,N] \}$,  $\textup{conv}\left\{\mathcal{X}_i\right\}_{i=1}^N:=$ $\{  \sum_{i=1}^{N}\lambda_i x_i \in \R^n \mid x_i \in \mathcal{X}_i, \lambda_i\in\left[0,1\right] \ \forall i \in \Z[1,N], \sum_{i=1}^N \lambda_i=1 \} $. Given two sets $ \mathcal{X}_{1}, \mathcal{X}_{2}$ and a vector $x_1\in\mathcal{X}_1$ we define the premetric  $\mu( \mathcal{X}_{1}, \mathcal{X}_{2}):=\sup_{x_1\in \mathcal{X}_{1}} \inf _{x_2\in   \mathcal{X}_{2}}\|x_1-x_2\|$ and $\mu(x_1,\mathcal{X}_2):=\inf _{x\in   \mathcal{X}_{2}}\|x_1-x\|$. Given $N$ vectors $v_i \in\R^n$, $i\in\Z\left[1,N\right]$, we denote $\left[v_1;\hdots;v_N\right]:=\left[v_1^\top \ldots v_N^\top\right]^\top\in\R^{Nn}$. 
Given $f,g:\mathbb{X}\subseteq \R^n\rightarrow\R$, $f(x)=\mathcal{O}(g(x))$ denotes that there exists $K>0$ such that $f(x)\le Kg(x)$ for all $x\in \mathbb{X}$.
\section{ Aggregative games over a network: Two motivating examples} \label{sec:motivation}
\subsection{Multi-dimentional constrained opinion dynamics in social networks with stubborn agents} \label{sec:mot_op} 
We first consider the problem of modeling  how ideas, innovations or behaviors spread in a social network of $N\in \N$ agents~\cite{montanari2010spread}. 
Generalizing the setting described in~\cite{ghaderi2014opinion}, we assume that each agent $i\in\Z\left[1,N\right]$ has a vector $x^i\in[0,1]^n$ of opinions regarding $n\in\N$ topics. Each component $x^i_s\in\left[0,1\right]$ represents the opinion of agent $i$  about topic $s\in\Z\left[1,n\right]$, where $0$ represents an extremely negative and $1$ an extremely positive opinion. We denote by $x^{i}_{(0)}\in\left[0,1\right]^n$ the initial opinion of agent $i$. Moreover, we specify the social network by the weighted adjacency matrix $P\in\R^{N\times N}$, where the element $P_{ij}\in\left[0,1\right]$ denotes the relevance of the opinion of agent $j$ to the decision of agent~$i$. In the following we assume without loss of generality that  $\sum_j P_{ij}=1$, for all $i\in\Z\left[1,N\right]$. 
To describe the opinion dynamics, we consider  a synchronous repetitive game where at every iteration $k$ each agent~$i$ communicates once with its neighbors  $\mathcal{N}^i:=\{j\in\Z[1,N]\mid P_{ij}>0\}$ and updates its opinion according to the  optimization problem
\begin{align} 
x^{i\,\star}\left(x^{\mathcal{N}^i}\right) := \argmin_{ x^i\in \R^n  }&\quad \!\!\! \sum_{j\neq i}^N (P_{ij}  \|x^i-x^j\|^2)\!+\theta_i\|x^i-x^{i}_{(0)}\|^2 \notag \\
\textup{s.t. }&  \quad x^i \in\mathcal{X}^i. \label{opinion} 
\end{align}
The cost function in~\eqref{opinion} comprises two terms: the first one models the influence of the neighbors to the new opinion of agent $i$, the second one models the ``stubbornness'' of agent~$i$ about its initial opinion. 
Additional constraints on the agents' opinions across the $n$ topics, as for example the fact that the opinions regarding two topics should not differ more than a given threshold, and on their stubbornness can be encoded via the  constraint set $\mathcal{X}^i\subseteq \left[0,1\right]^n $. The agents are assumed to be heterogeneous in the sense that the stubbornness parameter $\theta_i\ge0$, the constraint set $\mathcal{X}^i$ and the weights $\{P_{ij}\}_{j\ne i}^N$ may be different for every agent. Following the nomenclature used in~\cite{ghaderi2014opinion}, we refer to the agents for which $\mathcal{X}^i=\{x^i_{(0)}\}$, that is, agents that are not influenced by the neighbors, as \textit{fully stubborn}, to the agents for which $\theta_i=0$ and $\mathcal{X}^i =\left[0,1\right]^n$ as \textit{followers} and to all the remaining agents  as \textit{partially stubborn}. \\

 In the absence of constraints, the  solution to \eqref{opinion} for each topic decouples, hence we can consider $n=1$ without loss of generality. The solution  in this case is  given by
$$\textstyle x^{i\,\star}(x^{\mathcal{N}^i})=\frac{1}{1+\theta_i} \sum_{j\ne i} P_{ij}x^j +\frac{\theta_i}{1+\theta_i} x^{i}_{(0)},$$
which leads to  the BR dynamics $x^i_{(k+1)}=x^{i\,\star}(x^{\mathcal{N}^i}_{(k)})=\frac{1}{1+\theta_i} \sum_{j\ne i} P_{ij}x^j_{(k)} +\frac{\theta_i}{1+\theta_i} x^{i}_{(0)}$ which is a particular case of the standard Friedkin and Johnsen model \cite{friedkin1999social}, with parameters $\Lambda:=\textup{diag}\left(\frac{1}{1+\theta_1},\ldots,\frac{1}{1+\theta_N}\right)$ and $W:=P$.  For the case in which there is at least one partially stubborn agent  it is possible to show, with the same argument used in~\cite{ghaderi2014opinion}, that the  BR dynamics converge to a Nash equilibrium of the game in \eqref{opinion}, that is, to a configuration where no agent has incentive in modifying its opinion given the opinions of its neighbors. 
If on the other hand, all the agents are followers, one recovers the standard DeGroot model \cite{degroot1974reaching}, whose convergence properties have been exhaustively investigated using \textit{classical consensus theory}. We note that, in this case, the Nash equilibria coincide with the right eigenvectors of the matrix $P$ corresponding to the  eigenvalue $\lambda=1$.

As a corollary of the theory developed in this paper, we derive in Section~\ref{sec:opinion}  conditions under which the BR dynamics converge to a Nash equilibrium in the presence of  \textit{ stubborn agents and generic convex constraints}. Moreover, for the case in which the BR dynamics do not converge, we propose the use of alternative opinion-update rules that ensure convergence. To this end, we note that the cost function in~\eqref{opinion} can be rewritten, up to constant terms that do not depend on $x^i$, as
\begin{align} 
J^i(x^i,\sigma^i) := (1+\theta_i) \|x^i\|^2-2(\sigma^i+\theta_ix^{i}_{(0)})^\top x^i,\label{opinion2} \end{align}
where $\sigma^i=\sum_{j\neq i}^N P_{ij}x^{j}$ is the average opinion of the neighbors of agent $i$. Consequently, the game in~\eqref{opinion}  can be thought of as a game where each agent tries to minimize a cost function that depends on its own strategy $x^i$ and on a convex combination of the strategies of its neighbors $\sigma^i$, resulting in a NA game, as defined in Section~\ref{sec:problem_statement_1}.

\subsection{Demand-response methods and mean field control via local communications} \label{sec:mot_}
As second motivating example, we consider a population of $N\in \N$  loads whose electricity consumption 
$u^i=\left[ u_1^i, \ldots, u^i_{T} \right] \in\R^T$ over the horizon $\mathcal{T}=\Z\left[1,T\right]$ is scheduled according to the following demand-response scheme
\begin{align} \label{eq:demand_response}
 {u}^{i \, \star}( \bar \sigma ) :=   \argmin_{ {u} \in \R^T }  & \quad \sum_{t\in\mathcal{T}}\left( \rho_i\left\| {u_t} - {\hat{u}_t^i}   \right\|^2 + p(\bar \sigma_t) {u}_t\right) \\ 
\textup{s.t. } &\quad s_{t+1}=a^i s_t+\gamma^i u_t \quad \forall t\in\mathcal{T}\notag\\ 
&\quad {[s(u); u]}\in{(\mathcal{S}^i \times \mathcal{U}^i}) \cap \mathcal{C}^i ,\notag
\end{align}
where $s_t=s_t(u)$ is the state of the load at time $t$ (e.g., its temperature in case of heating ventilation air conditioning systems \cite{ma2014distributed}, as detailed in Section \ref{sec:hvac}, and thermostatically controlled loads \cite{grammatico15ecc} or its state of charge in case of plug-in electric vehicles \cite{ma:callaway:hiskens:13, parise:colombino:grammatico:lygeros:14,grammatico:parise:colombino:lygeros:14}), $s^i_1\in\R$ is the given initial state,  $a^i,\gamma^i\in\R\backslash\{0\}$ are  parameters modeling the dynamics  and the efficiency of  load $i$, $ \bar \sigma_t=\frac1N\sum_{i=1}^N {u}^{i }_t\in\R$ is the population aggregate energy demand at time~$t$, $\bar \sigma:=\left[ \bar \sigma_1; \ldots; \bar \sigma_{T} \right] \in\R^T$. The energy consumption~$u$ and state vector $s(u)$ are constrained by  
the personalized  sets $\mathcal{U}^i\subset\R^{T}$ and $\mathcal{S}^i\subset\R^{T}$, respectively, and by the coupling constraint set $\mathcal{C}^i\subset\R^{2T}$.
The first term in the cost function of~\eqref{eq:demand_response} models the curtailment cost that each agent  encounters for deviating from its nominal energy schedule $\hat{u}^i_t$, according to the Taguchi loss function~\cite{taguchi2005taguchi}, $\rho_i>0$ being a constant weighting parameter. The second term models the demand-response mechanism: the price that each agent has to pay for the required energy varies according to a price function $p(\bar \sigma_t)$ that is  increasing in the total energy demand  at time $t$. 
Since the price function depends on the average strategy $\bar \sigma$ of  \textit{all the players}, that is $\sigma^i=\bar \sigma$ for all agents, this is a \textit{deterministic quadratic mean field game}, as described in \cite{parise:colombino:grammatico:lygeros:14, bauso:pesenti:13, grammatico:parise:colombino:lygeros:14}. The associated \textit{mean field control problem} aims at designing a price vector $p(\bar \sigma)$ such that the set of optimal responses $\{{u}^{i \, \star}( \bar \sigma )\}_{i=1}^N$ possesses some desirable properties in terms of the MF game in \eqref{eq:demand_response}.
Approaches to solve this problem iteratively via a central coordinator 
 have been proposed in 
\cite{ma:callaway:hiskens:13,parise:colombino:grammatico:lygeros:14, grammatico:parise:colombino:lygeros:14,ma2014distributed}. 
Here we propose a distributed algorithm with guaranteed convergence to an equilibrium price signal, by using local communications only. We also investigate the relation of the resulting solution with the one obtained via  a central coordinator. 

\section{ Game setting and previous work} \label{sec:problem_statement}
\subsection{Constrained network aggregative games} \label{sec:problem_statement_1}

\begin{figure}[h!]
\begin{center}
\includegraphics[width=0.35\textwidth]{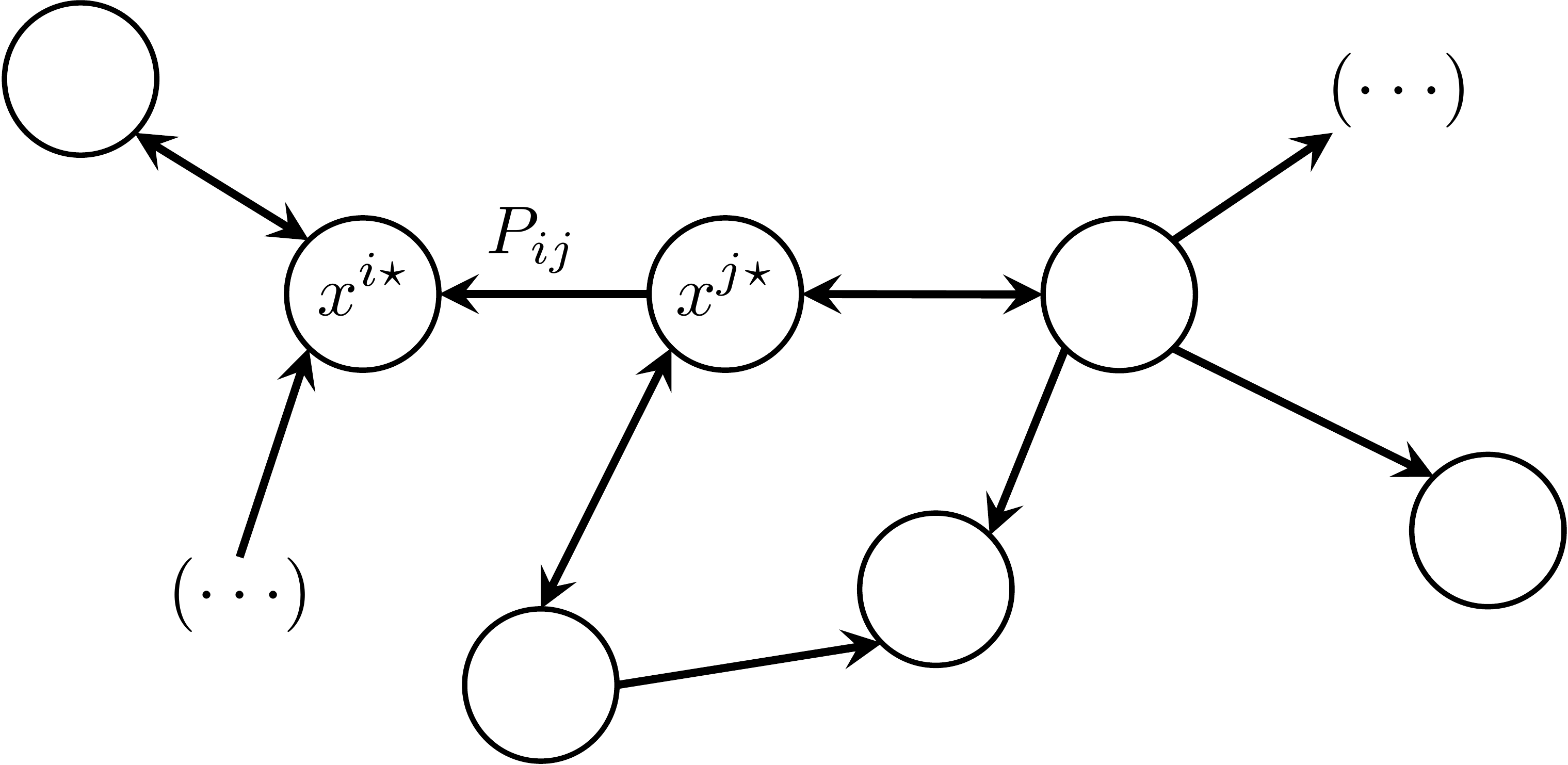}
\caption{Scheme of a network aggregative game.}
\label{fig:graph}
\end{center}
\end{figure}

Consider a population of $N \in \N$ heterogeneous agents,   where each agent $i \in \Z[1,N]$ controls a decision variable $x^i$, taking values in the  set $\mathcal{X}^i \subset \R^n$, and interacts with the other agents via a directed network (Figure \ref{fig:graph}). 
In the following, we specify the network by the weighted adjacency matrix $P\in\mathbb{R}^{N\times N}$, whose element  $P_{ij}\in\left[0,1\right]$ denotes the strength (or  relevance) of the communication from agent $j$ to agent $i$, $P_{ij}=0$ implying  no communication. The aim of agent $i$ is to minimize its individual deterministic cost 
$ \textstyle J^i\left(x^i, \sigma^i \right)$ 
that depends on its own strategy $x^i$ and its aggregation  $\sigma^i:= \sum_{j=1}^N  P_{ij}x^j$ with the strategies of its neighbors $\mathcal{N}^i:=\{j\neq i\mid P_{ij}>0\}$. 
Specifically, each agent $i \in \Z[1,N]$ aims at computing the \textit{Best Response} (BR) to the neighbors' aggregate state $\sigma^i$
\begin{align} \label{eq:nag}
x^{i}_{ \textup{br} }(\sigma^i ) &:= \argmin_{ x^i \in \mathcal{X}^i } J^i\left(x^i, \sigma^i \right)\\&\,= \argmin_{ x^i \in \mathcal{X}^i }  \textstyle J^i \left(x^i, P_{ii}x^i+\sum_{j\neq i}^N P_{ij}x^j \right). \notag
\end{align}
 Since the neighbors' aggregate state $\sigma^i$ is different for each agent and is specified by the network, we refer to this problem as a \textit{Network Aggregative} (NA) game.
In classical game theory, a set of strategies in which every agent is playing a best response to the other players' strategies is called Nash equilibrium. In the aggregative case, the concept is similar: the population is at a NA Nash equilibrium if each agent has no individual benefit in changing its strategy, given the \textit{aggregation} among the strategies of the neighbors.
\vspace{0.2cm}
\begin{definition}[Network aggregative Nash equilibrium] \label{def:Nash-equ-1}
Given $N$ cost functions $\{J^i: \R^n \times \R^n \rightarrow \R\}_{i=1}^N$ and a weighted adjacency matrix $P\in\R^{N\times N}$, a set of strategies $\left\{ \bar{x}^i\in\mathcal{X}^i\subseteq\R^n\right\}_{i=1}^{N}$ is a NA Nash equilibrium for~\eqref{eq:nag} if, for all $i \in \Z[1,N]$,
\begin{equation*}
J^i\left( \bar{x}^i, \textstyle \sum_{j=1}^{N} P_{ij} \bar{x}^j \right) =  \min_{ y \in \mathcal{X}^i } \textstyle J^i\left(y, P_{ii} y + \sum_{j\neq i}^{N}  P_{ij}  \bar{x}^j  \right).\hspace{0.2cm}\square
\end{equation*}
\end{definition}

One of the main challenges in analyzing aggregative games, and games in general, is to characterize the evolution of the players' strategies when the game is  repeated iteratively. For this problem to be well defined one has to specify the update rule used by each agent $i$, at iteration $k$, to select its updated optimal strategy $x^{i\,\star}_{(k+1)}$ in response to the strategies of its neighbors. The simplest dynamics are obtained when, at every iteration, all the players synchronously compute their BRs to the current strategies of the neighbors, that is
\begin{equation}\label{eq:brd}
x^{i\,\star}_{(k+1)}:=x^{i}_{\textup{br}}(\sigma^i_{(k)}),
\end{equation}
then communicate and  update the neighbors' aggregate state. This scheme is known as ``best response'' dynamics or ``myopic'' dynamics \cite{jensen:10}, since the agents do not take into account the past or future evolution of the game to update their strategies. One of the main aims of the paper is to derive conditions under which the BR dynamics converge to a Nash equilibrium of the original game as described in Definition~1.

In this paper, we restrict our attention to NA games satisfying the following  assumption.
\vspace{0.2cm}
\begin{assumption}[Game setting]\label{sa}
Each agent $i\in\Z[1,N]$ has a~convex and compact constraint set $\mathcal{X}^i\subset\R^n$ and quadratic cost 
\begin{equation}\label{eq:cost}
J^i(x^i,\sigma^i):=q_i {x^i}^\top Q x^i + 2 \left( C \sigma^i + c_i \right)^\top x^i,
\end{equation} 
where $x^i,\sigma^i  \in \R^n$,  $Q \succ 0$, $q_i>0$, $\C \in \R^{n \times n}$ and $\c_i \in \R^n$. The weighted adjacency matrix $P\in\R^{N\times N}$  is \textit{row stochastic}, that is, $\sum_{j=1}^N P_{ij}=1, \forall i\in\Z[1,N]$. \hfill{$\square$}
\end{assumption}
\vspace{0.2cm}
Note that the cost function in~\eqref{eq:cost} combines two different  terms: a quadratic cost specific to the  individual and an affine term associated with the neighbors' aggregate state $\sigma^i$. Under this assumption, the minimizer in \eqref{eq:nag} is unique.

\subsection{Aggregative games, deterministic mean field games and mean field control}  \label{sec:problem_statement_2}
If the network is fully connected, that is,  if $\sigma^i=\bar \sigma=\frac1N\sum_{j=1}^N x^j \in\R^n$ for all $i$, then the corresponding NA game is a particular case of aggregative game with aggregator  $\bar \sigma$. Aggregative games, in turn, have a strong connection with \textit{deterministic Mean Field} (MF) \textit{games}. The latter term is used in the literature to denote a game, played among a \textit{large} population of $N$ heterogeneous agents, where each agent is influenced by the statistical distribution of the strategies across the population. If each agent is influenced by the first moment of the distribution only, deterministic MF games coincide with aggregative games with aggregator $\bar \sigma$ and hence with NA games over fully connected networks. 
Mathematically, in all these games each agent $i\in\Z\left[1,N\right]$ aims at computing its BR
\begin{equation} \label{eq:mfg}
x^{i}_{\textup{br}}(\bar \sigma) := \argmin_{ x^i \in \mathcal{X}^i } J^i(x^i,\bar \sigma)
\end{equation}
 to the aggregate strategy $\bar \sigma$ of all the players. 

\vspace{0.2cm}
\begin{definition}[MF Nash equilibrium {\cite[Definition 1]{grammatico:parise:colombino:lygeros:14}}]   \label{def:Nash-equ-mf}
Given $N$ cost functions $\{J^i: \R^n \times \R^n \rightarrow \R\}_{i=1}^N$ and $\varepsilon > 0$, a set of strategies $\left\{ \bar{x}^i\in\mathcal{X}^i \subseteq \R^n\right\}_{i=1}^{N}$ is a MF $\varepsilon$-Nash equilibrium for~\eqref{eq:mfg} if, for all $i \in \Z[1,N]$,
\begin{equation} \label{eq:epsilon-Nash}
J^i\left( \bar{x}^i, \textstyle \frac{1}{N} \sum_{j=1}^{N}  \bar{x}^j \right) \leq  \min_{ y \in \mathcal{X}^i } \textstyle J^i\left(y, \frac{1}{N} \left(  y + \sum_{j\neq i}^{N}  \bar{x}^j \right) \right) + \varepsilon.
\end{equation}
It is a MF Nash equilibrium if~\eqref{eq:epsilon-Nash} holds with $\varepsilon=0$.
{\hfill $\square$}
\end{definition}
\vspace{0.2cm}


One of the objectives of \textit{mean field control} theory is to study  conditions under which it is possible to steer the population to such Nash equilibria, by  using global macroscopic incentives only, see for example~\cite{chen2014autonomous,ma:callaway:hiskens:13,grammatico:parise:colombino:lygeros:14}. 
Let
\begin{equation} \label{eq:opt_prob_mf}
x^{i\,\star}(z) := \argmin_{ x \in \mathcal{X}^i } J^i(x,z)
\end{equation}
be the optimal solution that agent $i$ would compute, according to the cost function $J^i$ given in~\eqref{eq:cost}, in response to a fixed global incentive $z$ and consider the aggregation mapping $\mathcal{A}: \R^n \rightarrow \left( \frac{1}{N} \sum_{i=1}^{N} \mathcal{X}^i \right) \subset \R^n$, defined as
\begin{equation} \label{eq:aggregator_single}
\textstyle \mathcal{A}(z) := \frac{1}{N} \sum_{i=1}^{N} x^{i \, \star}(z).
\end{equation}
For the quadratic cost function considered in \eqref{eq:cost}, it is proven in~\cite[Theorem 1]{grammatico:parise:colombino:lygeros:14} that $\mathcal{A}$ in \eqref{eq:aggregator_single} has a fixed point $\bar{z}=\mathcal{A}(\bar{z})$ and  the set of strategies $\left\{ x^{i \, \star}(\bar z)\in\mathcal{X}^i\right\}_{i=1}^{N}$ is an $\varepsilon_N$-Nash equilibrium for~\eqref{eq:mfg}, with $\varepsilon_N$ decreasing to zero with rate $\frac1N$ as the population size $N$ grows. To find such a fixed point, and hence steer the population to an $\varepsilon$-Nash equilibrium, one can use an iterative scheme among the agents and a central coordinator where, at every iteration step $k$, the central coordinator broadcasts a reference signal $z_{(k)}$ to the  population and each agent $i$ responds by computing its optimal strategy $x^{i\,\star}_{(k)}:=x^{i\,\star}(z_{(k)})$. The central coordinator then collects the average $\mathcal{A}(z_{(k)})$  of such strategies  and updates the reference signal according to $z_{(k+1)}=\Phi_k(z_{(k)},\mathcal{A}(z_{(k)}))$ ~\cite[ Algorithm~\ref{alg:br}]{grammatico:parise:colombino:lygeros:14}. The feedback mapping $\Phi_k(\cdot,\cdot)$ can be selected to guarantee convergence, depending on the  regularity properties of the aggregation mapping $\mathcal{A}$.  Suitable iterations and sufficient conditions on the matrices defining the cost in~\eqref{eq:cost} are given in~\cite[Corollary~1]{grammatico:parise:colombino:lygeros:14}. For example, if $\mathcal{A}$ is a contraction mapping (see Definition~\ref{def:CON} in Appendix A) then the Picard--Banach mapping $\Phi^{\textup{P--B}}(z_{(k)},\mathcal{A}(z_{(k)})):=\mathcal{A}(z_{(k)})$ ensures convergence to the unique fixed point of the aggregation mapping $\mathcal{A}$. 

Note that the centrally coordinated solution detailed above cannot be used to steer the agents to an equilibrium in NA games since, in this case, the signal $\sigma^i$ is different for every agent. On the contrary, we show below that the distributed algorithms derived for  NA games can be used to approximate the solution of MF games. For this purpose, we define one last class of games: NA games with multiple rounds of communications.


\subsection{Multiple-communication network aggregative games  solve the mean field control problem} \label{sec:problem_statement_3}
The solution to the MF control problem outlined in Section~\ref{sec:problem_statement_2} relies on the presence of a central coordinator that at every iteration computes the average signal $\mathcal{A}(z_{(k)})$ and broadcasts the reference signal $z_{(k+1)}$ to the population of agents.
 To solve the original MF problem by means of local communications only,
 we consider a scenario where each agent $i\in\Z\left[1,N\right]$ tries to recover the overall population average $\bar \sigma$ by communicating  $\nu\in\N$ times with its neighbors; we indicate the resulting local estimate by $\sigma^{ i}_{\nu}\in\mathbb{R}^n$. Formally, each agent $i\in\Z\left[1,N\right]$ approximates the optimization  problem  in \eqref{eq:mfg} with 
\begin{equation} \label{eq:nag_nu}
 x^{i}_{\textup{br}}(\sigma_{\nu}^i) := \argmin_{ x \in \mathcal{X}^i } J^i(x,\sigma_{\nu}^i),
\end{equation}
where $\sigma_{\nu}^i:=\sum_{j=1}^N P_{ij}^\nu x^j$ and $P_{ij}^\nu $ denotes the  element $(i,j)$ of $P^\nu$. 
Definition \ref{def:Nash-equ-1} on NA Nash equilibrium can be extended to the case of multiple communications as follows. 
\vspace{0.2cm}
\begin{definition}[ \!\!Multiple-communication\! NA \!Nash \!equilibrium] \label{def:Nash-equ}
Given $N$ cost functions $\{J^i: \R^n \times \R^n \rightarrow \R\}_{i=1}^N$, a weighted adjacency matrix $P\in\R^{N\times N}$ and a fixed number of communications $\nu\in\N$, a set of strategies $\left\{ \bar{x}^i\in\mathcal{X}^i\right\}_{i=1}^{N}$ is a NA Nash equilibrium for~\eqref{eq:nag_nu} if, for all $i \in \Z[1,N]$, it holds
\begin{equation*}
J^i\left( \bar{x}^i, \textstyle \sum_{j=1}^{N} P_{ij}^\nu \bar{x}^j \right) =  \min_{ y \in \mathcal{X}^i } \textstyle J^i\left(y, P_{ii}^\nu y + \sum_{j\neq i}^{N}  P_{ij}^\nu  \bar{x}^j  \right). \hspace{0.2cm} \square
\end{equation*}
\end{definition}
\vspace{0.2cm}

We show in the following how the update rules derived to guarantee convergence to a NA Nash equilibrium of the game in \eqref{eq:nag} can be applied also to the multiple-communication NA game in~\eqref{eq:nag_nu} to steer the agents' strategies to a MF $\varepsilon$-Nash equilibrium of the MF game in \eqref{eq:mfg}, in a distributed fashion, for large $N$ and $\nu$.

\section{On network aggregative and mean field\\ Nash equilibria}  \label{sec:fixed_point}
To address the tasks described in Sections \ref{sec:problem_statement_1} and \ref{sec:problem_statement_3}, we solve the following problems.
\begin{itemize}
\item \textit{Problem 1:} Given a single ($\nu=1$) or multiple ($\nu>1$) communication NA game, derive update rules and conditions under which the agents' strategies converge to a NA Nash equilibrium, as described in Definition~\ref{def:Nash-equ-1} and~\ref{def:Nash-equ}, respectively;
\item \textit{Problem 2:} Given a deterministic MF game as in \eqref{eq:mfg} over a population of size $N$ and a network with weighted adjacency matrix $P_N$, derive  update rules and conditions under which the agents' strategies converge, in a distributed fashion, to a MF $\varepsilon$-Nash equilibrium, as described in Definition~\ref{def:Nash-equ-mf}, with $\varepsilon$ decreasing as  $N$ and $\nu$ increase.
\end{itemize}
To solve these two problems, we start by defining the optimal response of an agent to a fixed signal $z^i\in\R^n$ as
\begin{equation} \label{eq:opt_prob}
 x^{i\,\star}(z^i) := \argmin_{ x^i \in \mathcal{X}^i } J^i(x^i,z^i).
\end{equation}
This mapping differs from the BR mapping since the second argument in the cost function $J^i(\cdot,\cdot)$  is not the neighbors' aggregate state $\sigma^i$ (that in general could depend on $x^i$), but is a fixed  signal $z^i$ (that does not depend on the optimization variable $x^i$). Moreover, the mapping in \eqref{eq:opt_prob} differs from  the one in \eqref{eq:opt_prob_mf} since the signal $z^i$ is possibly different for each agent.
Let
  $\ze:=\left[\zl^1 ; \hdots ;\zl^N\right]\in \mathbb{R}^{Nn}$ be
a vector of (possibly different) signals for each agent
and define the mapping $\xe:\mathbb{R}^{Nn}\rightarrow \mathcal{X}_{1\times N}$ as

\begin{equation}
\label{eq:x_extended}
\xe(\ze):=\left[x^{1\, \star}(\zl^1); \hdots ; x^{N\, \star}(\zl^N)\right]\in \mathbb{R}^{Nn},
\end{equation}
whose sub-vectors are the optimal strategies computed by each agent $i$ according to the local signal~$z^i$.
The mapping $\xe$ in~\eqref{eq:x_extended} can be used to define an \textit{extended aggregation mapping} $\mn$ that, given a  vector $\ze$, returns the updated estimates of the population average, after one optimization and $\nu$ communication steps.  Formally, $\mn:\mathbb{R}^{Nn}\rightarrow (\Pn\otimes I_n)\mathcal{X}_{1\times N}\subset \mathbb{R}^{Nn} $ is defined as
\begin{align}\label{eq:mn}
 \mn(\ze)&:=\left[\begin{array}{c}\mathcal{A}_{\nu}^1(\ze) \\\vdots \\\mathcal{A}_{\nu}^N(\ze)\end{array}\right]:=\left[\begin{array}{c}\sum_{j=1}^N P_{1j}^\nu x^{j\,\star}(z^j) \\\vdots \\\sum_{j=1}^N P_{Nj}^\nu x^{j\,\star}(z^j)\end{array}\right]\\&\,\,=(\Pn\otimes I_n)\xe(\ze)=:\Pen \xe(\ze). \notag
 \end{align}

\subsection{Problem 1: Nash equilibria of single and multiple-communication network aggregative games  }

In the following theorem we show that the fixed points of the aggregation mapping $\mn$ can be used to find a  Nash equilibrium of the NA game with $\nu$ communications, for any population size $N$, under the following  assumption on the network structure.

 \vspace{0.2cm}
\begin{assumption}[Graph property] \label{a3}
 For the considered number $\nu\in\N$ of communication steps  and the given population size~$N$, the weighted adjacency matrix $P$  satisfies $P^\nu_{ii}=0$ for all   $i\in \Z[1,N].$
{\hfill $\square$}
\end{assumption}
\vspace{0.2cm}
\begin{remark} Assumption \ref{a3} is equivalent to the absence of  cycles of length  $\nu$ in   the graph associated with $P$ \cite[Theorem~4.1]{Biggs:2007}.   In the case of single-communication NA games, that is if $\nu=1$ as defined in Section~\ref{sec:problem_statement_1}, Assumption~\ref{a3} is  equivalent to the absence of self-loops, which means that each agent plays against the average of its neighbors, itself excluded.  If $P_{ii}/q_i$ is the same for each agent~$i\in\Z\left[1,N\right]$ this assumption can be introduced without loss of generality by redefining the $Q$ matrix of the cost function $J^i$ in \eqref{eq:cost}.   Note that the absence of self-loops is necessary but not sufficient  when $\nu>1$.
\label{loop}
\end{remark}
\vspace{0.2cm}
Assumption~\ref{a3} guarantees that the neighbors' aggregate state $\sigma^i_\nu$ computed by  each agent $i$ does not depend on its own strategy $x^i$. As a consequence, the BR mapping~\eqref{eq:nag_nu} coincides with the optimal response~\eqref{eq:opt_prob} to the signal $z^i:=\sigma^i_\nu=\sum_{j=1}^N P^\nu_{ij}x^j=\sum_{j\neq i}^N P^\nu_{ij}x^j$. Therefore, if each agent is playing the BR  to the $i$-th sub-vector $\bar{z}^i$ of a fixed  point $\bar{\ze}=\left[\bar{z}^1;\ldots,\bar{z}^N\right]$ of the aggregation mapping~$\mn$, then the new vector of neighbors' aggregate states coincides with $\bar{\ze}$. This observation can be formalized in the following theorem.
\vspace{0.2cm}
\begin{theorem}\label{thm:problem 2}
Under Assumption~\ref{sa} the mapping $\mn$ in~\eqref{eq:mn} admits at least one fixed point $\bar{\ze}=\mn(\bar{\ze})$.
If also Assumption~\ref{a3} holds and $\bar{\ze}$ is a fixed point of $\mn$, then the set of strategies 
$\left\{ x^{i \, \star}\left( \bar{z}^i \right)\right\}_{i=1}^{N}$, 
with $x^{i \, \star}$ as in~\eqref{eq:opt_prob} for all $i \in \Z[1,N]$, is a NA Nash equilibrium for~\eqref{eq:nag_nu}.
{\hfill $\square$}
\end{theorem}
\vspace{0.2cm}

Note that Theorem~\ref{thm:problem 2} implicitly  ensures the existence of at least one NA Nash equilibrium for all NA games satisfying Assumptions~\ref{sa} and \ref{a3}. 
Moreover,  under these assumptions, if the vector $\left[\sigma^1_{\nu\,(k)};\ldots,\sigma^N_{\nu\,(k)}\right]$ of neighbors' aggregate states defined by the BR dynamics in \eqref{eq:brd} tends, as $k\rightarrow\infty$, to a fixed point of the mapping  $\mn$, then the agents' strategies converge to a NA Nash equilibrium. The derivation of conditions under which this happens and of alternative update rules is postponed to Section~\ref{sec:quest}. 
Finally, we note that MF games are NA games over a complete network with the same self-weight $P_{ii}=1/N$ for all agents. If $q_i=q$ for all $i$,  these games can be reformulated in a form compatible with Assumption \ref{a3}, as suggested in Remark~\ref{loop}. Theorem~\ref{thm:problem 2} then guarantees that a MF Nash equilibrium exists (as shown also in \cite[Proposition 1]{grammatico:parise:colombino:lygeros:14}) and  can be obtained in a distributed fashion. The drawback is that, to this end, communications over a complete network are required. Since in most cases this is not desirable  in the following section we derive  conditions under which almost MF Nash equilibria can be recovered using local communications only.

\subsection{Problem 2: Distributed solution to the mean field control problem}
To address \textit{Problem 2}, we start by analyzing the relation between the extended aggregation mapping $\mn$ in~\eqref{eq:mn}, obtained via $\nu$ communications over the network $P$,  and the aggregation mapping $\m:\mathbb{R}^{Nn}\rightarrow (\ones\otimes I_n)\mathcal{X}_{1\times N}$, 
\begin{equation}\label{eq:m}
\textstyle \m(\ze):=\left(\ones \otimes I_n\right)\xe(\ze)=:\Ie\, \xe(\ze ),
\end{equation}
that arises out of a complete network.
We assume the following uniform bounds on the constraint sets $\mathcal{X}^i$ and on $q_i$.

\vspace{0.2cm}
\begin{assumption}[Compactness] \label{a1}
There exists  a compact set $\mathcal{X} \subset \R^n$ and $\underline q,\bar q>0$ 
such that, for all population sizes~$N$, $\textup{conv}\left(\left\{\mathcal{X}^i\right\}_{i=1}^N\right) \subseteq \mathcal{X}$ and $\underline q\le q_i\le \bar q$ for all $i\in\Z\left[1,N\right]$. 
{\hfill $\square$}
\end{assumption}\

For the game in \eqref{eq:nag_nu} to approximate the MF game in \eqref{eq:mfg}, we need to ensure that  $\lim_{\nu \rightarrow \infty} \sigma_{\nu}^i = \bar \sigma=\frac{1}{N}\sum_{j=1}^N x^j$, that is the agents  asymptotically reach consensus on the population average. To this aim we introduce the following assumption.

\vspace{0.2cm}
\begin{assumption}[Asymptotic average consensus] \label{a2}
 For all population sizes~$N$, the weighted adjacency matrix $P=P_N$ satisfies $\lim_{\nu\rightarrow \infty}P^\nu=\ones.$ Equivalently, $P$ is primitive and doubly stochastic.
{\hfill $\square$}
\end{assumption}
\vspace{0.1cm}

\begin{remark}
Let us prove the equivalence of the two statements in Assumption~\ref{a2}.
If $\lim_{\nu\rightarrow \infty}P^\nu=\ones$ then there exists  $h\in\N$ such that $\max_{i,j} | P_{i,j}^{h} - 1/N | < 1/N$. Consequently, there exists an $h>0$ such that $P^h$ is element-wise strictly positive, which means that  the matrix $P$ is primitive.  For a primitive and row stochastic matrix, by Perron--Frobenius theorem~\cite[Lemma 4]{olfati2007consensus},  $\lim_{\nu\rightarrow \infty}P^\nu=\one w^\top$, where $w^\top$ is the left eigenvector of $P$ relative to the dominant eigenvalue $1$ and normalized so that $w^\top\one=1$. It follows that $\lim_{\nu\rightarrow \infty}P^\nu=\ones \Leftrightarrow w^\top=\frac{1}{N} {\one}^\top  \Leftrightarrow  {\one}^\top P={\one}^\top \Leftrightarrow  P$ is doubly stochastic.
\end{remark}
\vspace{0.1cm}
Note that  Assumptions~\ref{a3} and~\ref{a2}  both regard  the matrix $P^\nu$ but  are used to address two different problems and they are conceptually different. Assumption~\ref{a3} imposes $P^\nu_{ii}=0$ for the \textit{fixed} value of $\nu$ considered in the NA game under analysis. 
 Assumption~\ref{a2}, on the other hand, is needed when one wants to use multiple-communication NA games to approximate the solution to a  MF game. Consequently, what matters is the \textit{asymptotic} behavior of $P^\nu_{ii}$, which should tend to the contribution that each agent has in a MF game, i.e. $1/N$.

 \vspace{0.2cm}

\begin{lemma}
Under Assumption ~\ref{sa}, the mappings $\m$ in~\eqref{eq:m} and $\mn$ in~\eqref{eq:mn}, for all $\nu \in \N$, are continuous and have at least one fixed point. 
 If  additionally Assumption~\ref{a2} holds, then
$\lim_{\nu \to \infty} \sup_{\ze \in  \R^{Nn}} \|\mn(\ze)-\m(\ze) \| 
= 0.$
{\hfill $\square$}
\label{lemma:SA}
\end{lemma}
\vspace{0.2cm}
We next  show that, by choosing $\nu$ large enough, any fixed point of $\mn$ is arbitrarily close to a fixed point of $\m$. 

\vspace{0.2cm}
\begin{lemma}[Fixed point sets]\label{th:continuityFixedPoints}
Let the sets of fixed points of $\m$ in~\eqref{eq:m}  and $\mn$ in~\eqref{eq:mn} be
$\mathcal{F}:=\left\{\ze \in \R^{Nn} \mid \ze=\m(\ze) \right\}$ and $\mathcal{F}_\nu:=\left\{\ze \in \R^{Nn} \mid \ze=\mn(\ze) \right\}$, respectively.
 If Assumptions~\ref{sa} and~\ref{a2} hold, then for all
$\epsilon >0 $ 
there exists $\bar{\nu}>0$ such that $\mu(\mathcal{F}_{\nu},\mathcal{F})\le\epsilon$ for all $\nu\ge \bar{\nu}.$ {\hfill $\square$}
\end{lemma}
\vspace{0.2cm}

We are now ready to state our main theorem regarding the solution to \textit{Problem 2}.
\vspace{0.2cm}
\begin{theorem}\label{thm:problem 1}
Suppose that Assumptions~\ref{sa},~\ref{a1} and~\ref{a2} hold. For all $\varepsilon>0$ there exists $\bar N$ such that, for all $N>\bar N$, there exists $\bar \nu>0$ such that, for all $\nu \ge \bar \nu$, if $\bar{\ze}$ is a fixed point of $\mn$ in~\eqref{eq:mn}, then the set of strategies 
$\left\{ x^{i \, \star}\left( \bar{z}^i \right)\right\}_{i=1}^{N}$, 
with $x^{i \, \star}$ as in~\eqref{eq:opt_prob} $\forall \, i \in \Z[1,N]$, is a MF $\varepsilon$-Nash equilibrium for~\eqref{eq:mfg}.
{\hfill $\square$}
\end{theorem}
\vspace{0.2cm}

We emphasize that, contrary to the case of NA games, where the fixed point of $\mn$ leads, under Assumption~\ref{a3}, to a NA Nash equilibrium for finite $\nu$ and finite $N$, in the case of MF games a MF Nash equilibrium is recovered only asymptotically. This is due to the fact that, first, the agents are required to almost reach consensus on the population average (hence $\nu$ should be large enough) and second, the population average  depends on the strategy of agent $i$ with contribution proportional to $1/N$ (hence $N$ should be large enough).
However,  for any desired $\varepsilon>0$, the proof of Theorem \ref{thm:problem 1} allows one to derive lower bounds on $N$ and $\nu$ in order to guarantee that $\left\{ x^{i \, \star}\left( \bar{z}^i \right)\right\}_{i=1}^{N}$ is a MF  $\varepsilon$-Nash equilibrium. 
We note that the minimum number of required communications $\bar\nu$ can be computed in a distributed fashion (Appendix C) and depends on $\varepsilon$ but also on the population size $N$ and on the network $P$.  In case of symmetric networks this dependence can be further specified in terms of the spectral properties of $P$.
\vspace{0.2cm}
\begin{assumption}[Spectral properties]\label{a2b}
For all population sizes~$N$ the   weighted adjacency matrix $P\!=\!P_N$~is symmetric. There exists $\mu\in\left[0,1\right)$ such $\mu_N:=\max_{\lambda\in\Lambda(P_N)\backslash\{1\}}|\lambda|\le \mu$ for all $N$. {\hfill $\square$}

\end{assumption}
\vspace{0.2cm}

\begin{corollary}\label{cor:problem 1 rate}
Suppose that Assumptions~\ref{sa},~\ref{a1},~\ref{a2} and~\ref{a2b} hold. If $\bar{\ze}$ is a fixed point of $\mn$ in~\eqref{eq:mn}, then the set of strategies 
$\left\{ x^{i \, \star}\left( \bar{z}^i \right)\right\}_{i=1}^{N}$, 
with $x^{i \, \star}$ as in~\eqref{eq:opt_prob} for all $i \in \Z[1,N]$, is a MF $\varepsilon$-Nash equilibrium for~\eqref{eq:mfg}, with $\varepsilon=\mathcal{O}(\frac1N+\sqrt{N}\mu^\nu).$
{\hfill $\square$}
\end{corollary}
\vspace{0.2cm}
In other words, Assumption~\ref{a2b} allows us to derive an upper bound on $\varepsilon$ that is composed by two terms: one that decreases linearly in $1/N$, as for the MF control solution with central coordinator \cite[Remark 3]{grammatico:parise:colombino:lygeros:14}, and one that, for any fixed $N$, decreases exponentially fast on the number of communication steps $\nu$.  According to this bound, the number of  communication steps $\bar\nu$ required to achieve a fixed tolerance $\varepsilon$ increases at most logarithmically in the population size $N$. \\
Assumption~\ref{a2b} is satisfied, e.g., by the degree-normalized adjacency matrices of any family of $d$-regular undirected $\epsilon$-expander graphs \cite[Definition 2.2 and Example 2.2]{hoory2006expander}. In fact $\mu_N\le 1-\frac{\epsilon^2}{2d^2}=:\mu$ \cite[Theorem 2.4]{hoory2006expander}, \cite[Theorem 2]{sinclair1992improved}.

\section{Iterative schemes and fixed point convergence }  \label{sec:quest}
We now return to update schemes that, under different conditions, will ensure convergence to a fixed point of  $\mn$.

\begin{algorithm} 
\caption{Memoryless update rule (BR)}
\label{alg:br}

\textbf{Initialization}. Set $k \leftarrow1$,  choose an initial reference $z^i_{(1)}$ for every agent $i$.

\vspace{0.01cm}

\textbf{Iterate until convergence}. 

\vspace{0.01cm}

Optimization step: each agent $i$ computes its optimal strategy with respect to the reference $z^i$

\vspace{0.01cm}

$\quad$ $\quad$ $\displaystyle x^{i \, \star} \leftarrow \argmin_{x \in \mathcal{X}^i} J^i(x,z^i_{(k)})$;

Communication step: each agent $i$ updates its neighbors' state by communicating $\nu$ times

$\quad$$\quad$ $ \sigma^i_{0} \leftarrow  x^{i \star}$;

\vspace{0.01cm}

$\quad$$\quad$ {\small\texttt{for  $s=0$ to $s=\nu-1$ do}}

\vspace{0.01cm}

$\quad$$\quad$ $\quad$ $ \sigma^i_{s+1} \leftarrow  \sum_{j=1}^{N} P_{ij} \sigma^j_{s}$;

\vspace{0.01cm}

$\quad$$\quad$ {\small\texttt{end }}

  and updates the reference

\vspace{0.01cm}

$\quad$$\quad$ $ z^i_{(k+1)} \leftarrow  \sigma^i_{\nu}$.

\vspace{0.01cm}

$\quad$ $k \leftarrow k+1$;
\end{algorithm}
The simplest update rule (Algorithm~1)  is the one where each agent  selects as strategy its optimal response to the current neighbors' aggregate state, that is  $z^i_{(k)}=\sigma^i_{\nu\, (k)}:=P^\nu_{ii}x^{i\,\star}_{(k)}+\sum_{j\neq i}^NP_{ij}^\nu x^{j\,\star}_{(k)}$. Mathematically, we can describe an iteration of this algorithm as $\ze_{(k+1)}=\mn(\ze_{(k)})$. Under Assumption~\ref{a3}, the optimal response coincides with the BR, hence Algorithm~\ref{alg:br} coincides with  the BR dynamics.
\begin{algorithm}
\caption{Update rule with memory}
\label{alg:mem}

\textbf{Initialization}. Set $k \leftarrow1$, choose an initial reference $z^i_{(1)}$ for every agent $i$, a feedback mapping $\left\{ \Phi_k \right\}_{k=1}^{\infty}$ and $\nu_1,\nu_2\in\Z_{\ge0}$, $\nu_1+\nu_2=\nu$.

\vspace{0.01cm}

\textbf{Iterate until convergence}.

\vspace{0.01cm}

Communication step 1:  each agent $i$ updates its estimate  by communicating $\nu_1$ times

$\quad$$\quad$  $ \mathcal{A}^i_{0,0} \leftarrow  z_{(k)}^i$;

\vspace{0.01cm}

$\quad$$\quad$ {\small\texttt{for  $s=0$ to $s=\nu_1-1$ do}}

\vspace{0.01cm}

$\quad$$\quad$ $\quad$ $\mathcal{A}^i_{s+1,0} \leftarrow  \sum_{j=1}^{N} P_{ij} \mathcal{A}^j_{s,0}$;

\vspace{0.01cm}

$\quad$$\quad$ {\small\texttt{end}}
 
 \vspace{0.01cm}

\vspace{0.01cm}

Optimization step: each agent $i$ computes its optimal strategy with respect to the current estimate $ \mathcal{A}^i_{\nu_1,0}$

\vspace{0.01cm}

$\quad$ $\quad$ $\displaystyle x^{i \, \star} \leftarrow \argmin_{x \in \mathcal{X}^i} J^i(x, \mathcal{A}^i_{\nu_1,0})$;

\vspace{0.1cm}

Communication step 2: each agent $i$ updates its estimate by communicating $\nu_2$ times

$\quad$$\quad$ $ \mathcal{A}^i_{\nu_1,0} \leftarrow  x^{i \star}$;

\vspace{0.01cm}

$\quad$$\quad${\small\texttt{ for  $s=0$ to $s=\nu_2-1$ do}}

\vspace{0.01cm}

$\quad$$\quad$ $\quad$ $ \mathcal{A}^i_{\nu_1,s+1} \leftarrow  \sum_{j=1}^{N} P_{ij}  \mathcal{A}^j_{\nu_1,s}$;

\vspace{0.01cm}

$\quad$$\quad$ {\small\texttt{end}}
 
 \vspace{0.01cm}

  and updates the reference

\vspace{0.01cm}

$\quad$$\quad$ $ z_{(k+1)}^i \leftarrow  \Phi_k\left( z_{(k)}^i, \mathcal{A}^i_{\nu_1,\nu_2}\right) $.

\vspace{0.01cm}

$\quad$ $k \leftarrow k+1$;

\end{algorithm}

To allow convergence under more general conditions, we consider extensions of Algorithm~1 where at every step $k$ each agent~$i$  
computes its optimal response to a filtered version of  $\sigma^i_{\nu\,(k)}$, instead of myopically computing the BR.  Specifically, we assume that the agents perform $\nu_1\in\left[0,\nu\right]$ communication steps before updating their strategies and $\nu_2=\nu-\nu_1$ communications afterwards.  Let $\ze_{(k)}$ be the vector of local signals at the beginning of step $k$.  
After the first $\nu_1$ rounds of communications agent $i$ obtains the  averaged value $\mathcal{A}^i_{\nu_1,0}(\ze_{(k)}):=\sum_j P^{\nu_1}_{ij}z_{(k)}^j$ (where the two subscripts of $\mathcal{A}^i_{\nu_1,0}$ refer to the number of communication steps before and after the optimization step). Based on this, it computes its optimal strategy $x^{i \, \star}_{(k)}:=x^{i \, \star}(\mathcal{A}^i_{\nu_1,0}(\ze_{(k)}))$, and then 
communicates the remaining $\nu_2$ times with its neighbors, obtaining $\mathcal{A}^i_{\nu_1,\nu_2}(\ze_{(k)}):=\sum_j P^{\nu_2}_{ij} x^{j\,\star}(\mathcal{A}^j_{\nu_1,0}(\ze_{(k)}))$. Finally, it updates the local signal $z^i$ using a feedback mapping $\Phi_k$ that filters the current  aggregate state $\mathcal{A}^{i}_{\nu_1,\nu_2}(\ze_{(k)})$ with the previous local signal $z^i_{(k)}$, that is $ z_{(k+1)}^i = \Phi_{k}\left( z_{(k)}^i, \mathcal{A}^i_{\nu_1,\nu_2}(\ze_{(k)})\right) $.
Since the agents use information about the previous signal, we refer to Algorithm~\ref{alg:mem} as update rule  with memory.

Let $\bold{\Phi}:\R^{Nn}\times \R^{Nn} \rightarrow\R^{Nn}$ be the extended feedback mapping, defined as
\begin{equation}\label{eq:phi}\bold{\Phi}(\ze,\m_{\nu_1,\nu_2}):=\left[\Phi(\zl^1,\mathcal{A}^1_{\nu_1,\nu_2}); \hdots;  \Phi(\zl^N,\mathcal{A}^N_{\nu_1,\nu_2})\right],\end{equation}
 and note that, by construction, it can be computed in a  distributed fashion. An iteration of Algorithm~\ref{alg:mem} can then be described by
$\ze_{(k+1)} = \bold{\Phi}( \ze_{(k)}, \m_{\nu_1,\nu_2} \left( \ze_{(k)} \right) )$.  As candidate feedback mappings, we consider different classes of fixed point iterations from operator theory. The simplest feedback mapping is the Picard--Banach iteration~\cite[Theorem 2.1]{berinde}
\begin{equation} \label{eq:Picard-Banach}
\ze_{(k+1)}\! = \bold{\Phi}^{ \text{P--B} }( \ze_{(k)}, \m_{\nu_1,\nu_2} \left( \ze_{(k)} \right) )\!:=\m_{\nu_1,\nu_2}\left( \ze_{(k)} \right).
\end{equation}
Using this mapping and setting $(\nu_1,\nu_2)=(0,\nu)$ it is easy to see that Algorithm \ref{alg:br} is a particular case of Algorithm~\ref{alg:mem}; from now on we therefore restrict our attention to the latter algorithm. More general fixed point iterations are the Krasnoselskij iteration~\cite[Theorem 3.2]{berinde}
\begin{equation} \label{eq:Krasnoselskij}
\bold{\Phi}^{ \text{K} }(\ze_{(k)}, \m_{\nu_1,\nu_2}\left( \ze_{(k)} \right)):=(1 - \lambda)\ze_{(k)} + \lambda \m_{\nu_1,\nu_2} \left( \ze_{(k)} \right) 
\end{equation}
with $\lambda \in (0,1)$, and the step-dependent Mann iteration~\cite[Definition 4.1]{berinde}
\begin{equation} \label{eq:Mann}
 \bold{\Phi}_k^{ \text{M} }( \ze_{(k)}, \m_{\nu_1,\nu_2}(\ze_{(k)}) ) := (1-\alpha_k) \ze_{(k)} + \alpha_k \m_{\nu_1,\nu_2}(\ze_{(k)}), 
\end{equation}
where the sequence $\left(\alpha_k\right)_{k=1}^{\infty}$ is such that  $\alpha_k \in (0,1) \ \forall k \geq 0$, $\lim_{k \rightarrow \infty} \alpha_k = 0$ and $\sum_{k=1}^{\infty} \alpha_k = \infty $ (e.g., $\alpha_k=1/k$).

The following result provides  conditions on the cost functions and on the network structure under which the sequence of vectors used to compute the optimal responses,
\begin{equation}\label{ak}
\left(\left[ \mathcal{A}^1_{\nu_1,0}(\ze_{(k)}); \, \hdots ;\, \mathcal{A}^N_{\nu_1,0}(\ze_{(k)}) \right] \right)_{k=1}^\infty,
\end{equation} 
converges as $k$ tends to infinity to a fixed point of the aggregation mapping $\mn$ in~\eqref{eq:mn}. As a consequence,  $\{x^{i \, \star}_{(k)}\}_{i=1}^N$ converges to the desired Nash equilibrium configuration, according to Theorem~\ref{thm:problem 2} and Theorem~\ref{thm:problem 1}.
Let
\begin{equation} \label{eq:con} \textstyle
M_i:=\left[
\begin{matrix}
q_iQ  &  - C \\
- C^\top & q_iQ 
\end{matrix} 
\right], \quad i\in\Z\left[1,N\right].
\end{equation}

\begin{theorem} \label{convergence}
Under Assumption ~\ref{sa}, the following iterations and conditions guarantee that the sequence in~\eqref{ak} converges, for any initial vector $\ze_0 \in \R^{Nn}$, to a fixed point of $\mn$ in~\eqref{eq:mn}.

\vspace{0.2cm}

\hspace{-0.5 cm}\begin{tabular}{llcccc}
&& eq.& $(\nu_1,\nu_2)$ & cost ($\forall i$) & network \\ \toprule
1. & $\Phi^{ \text{P--B} }$ & \eqref{eq:Picard-Banach} &  $(0,\nu)$ & $M_i\succ0$  & $\|P\|\le1$ \\
2. & $\Phi^{ \text{K} }$ & \eqref{eq:Krasnoselskij} &  $(0,\nu)$  & $M_i\succcurlyeq0$ & $\|P\|\le1$ \\
3. & $\Phi^{ \text{P--B} }$ & \eqref{eq:Picard-Banach} & $(\nu/2,\nu/2) $  & $-q_iQ \preccurlyeq \C \prec 0$ & $P=\!P^\top\!$ \\
4. & $\Phi_k^{ \text{M} }$ & \eqref{eq:Mann} &  $(\nu/2,\nu/2) $  & $C \succ 0 $ & $P=P^\top$  \\\toprule
\end{tabular}\\[0.1cm]
The mapping $\mn$ has a unique fixed point in case 1. {\hfill $\square$}

\end{theorem}
\vspace{0.2cm}
The assumption $P=P^\top$ corresponds to having an undirected graph with symmetric weights. The assumption $\|P\|\le1$ is more involved. In general, the 2-norm  coincides with the largest singular value of the matrix $P$. In case of an undirected symmetric graph, the singular values coincide with the magnitude of the  eigenvalues, hence the $2$-norm results in the spectral radius of the matrix $P$, which under the assumption of row-stochasticity is $1$. Hence the condition $\|P\|\le1 $ is always satisfied for  symmetric graphs. In the case of directed graphs one can use H\"{o}lder's inequality $\|P\|\le \sqrt{\|P\|_1\|P\|_\infty}$ to prove that any doubly-stochastic matrix $P$ satisfies the assumption.

   It follows from the proof of Theorem~\ref{convergence}  that whenever $\Phi^{ \text{P--B} }$ is applicable so are $\Phi^{ \text{K} }$ and  $\Phi^{ \text{M} }$, and whenever $\Phi^{ \text{K} }$ is applicable so is $\Phi^{ \text{M} }$.
Which one among  the possible choices of  feedback mappings provides the best convergence performance is in general problem dependent; known convergence rates are available for $\Phi^{ \text{P--B} }$ \cite[Theorem 2.1]{berinde}.

\section{Network applications}  \label{sec:applications}

\subsection{Multi-dimentional constrained opinion dynamics in social networks with stubborn agents}\label{sec:opinion}
The constrained opinion dynamics problem  introduced in Section \ref{sec:mot_op} can be rewritten as a single-communication ($\nu=1$) NA game, as defined in~\eqref{eq:nag}, by using the cost function~\eqref{eq:cost} with $q_i:=(1+\theta_i), Q=I_n, C=-I_n$, $c=-\theta_ix^{i}_{(0)}$ and constraint set $\mathcal{X}^i$. Since $\sigma^i$ does not depend on $x^i$, $P_{ii}=0$ for every $i$ and Assumption~\ref{a3} is satisfied. Consequently, by Theorem~\ref{thm:problem 2}, if $\bar{\ze}$ is a fixed point of the mapping $\boldsymbol{\mathcal{A}}_{1}$ in~\eqref{eq:mn}, then the set of strategies $x^{i\,\star}(\bar{z}^i)$ is a NA Nash equilibrium. The following result shows that if $\theta_i>0$ for all $i\in\Z\left[1,N\right]$ and $\|P\|\le1$, the BR dynamics converge to the unique fixed point of $\boldsymbol{\mathcal{A}}_{1}$. If some \textit{followers} are present in the population, then a Nash equilibrium configuration can be reached using the update rule with local memory.
\vspace{0.2cm}

\begin{corollary}\label{cor:opinion}
Suppose that Assumption ~\ref{sa} holds.
The following iterations and conditions guarantee convergence of the opinions computed according to Algorithm~\ref{alg:mem}, from any initial configuration, to a NA Nash equilibrium for~\eqref{opinion}.

\begin{center}\begin{tabular}{cllccccr}
&&& eq.& $(\nu_1,\nu_2)$ & cost ($\forall i$) & network \\ \cmidrule[.75pt](lr){2-7}
&1. & $\Phi^{ \text{P--B} }$ &\eqref{eq:Picard-Banach} &  $(0,1)$  &  $\theta_i>0$ & $\|P\|\le1$;& \\
&2. & $\Phi^{ \text{K} }$ & \eqref{eq:Krasnoselskij} &  $(0,1)$ &   &$\|P\|\le1$.  & \hspace{0.3cm}$\square$\\ \cmidrule[.75pt](lr){2-7}
\end{tabular}\end{center}
\end{corollary}

\vspace{0.2cm}

Note that Algorithm~\ref{alg:mem} associated with the feedback mapping $\Phi^{ \text{P--B} }$ and $(\nu_1,\nu_2)=(0,1)$  corresponds to the BR dynamics (Algorithm~\ref{alg:br}) that are typically studied in the  literature of coordination games. The use of the feedback mapping $\Phi^{ \text{K} }$, on the other hand, extends the previous setting to the case of agents with memory.  This additional feature allows one to recover a Nash equilibrium in the more general case of a population possibly containing both {stubborn agents} and {followers}. Note that the conditions of Corollary~\ref{cor:opinion} are only sufficient, therefore there might be cases where the BR dynamics converge even in the presence of followers. It is however possible to find cases where the BR dynamics do not converge;  an example is a population of  followers communicating through a directed ring network.   

\begin{figure}
\begin{center}
\includegraphics[width=0.4\textwidth]{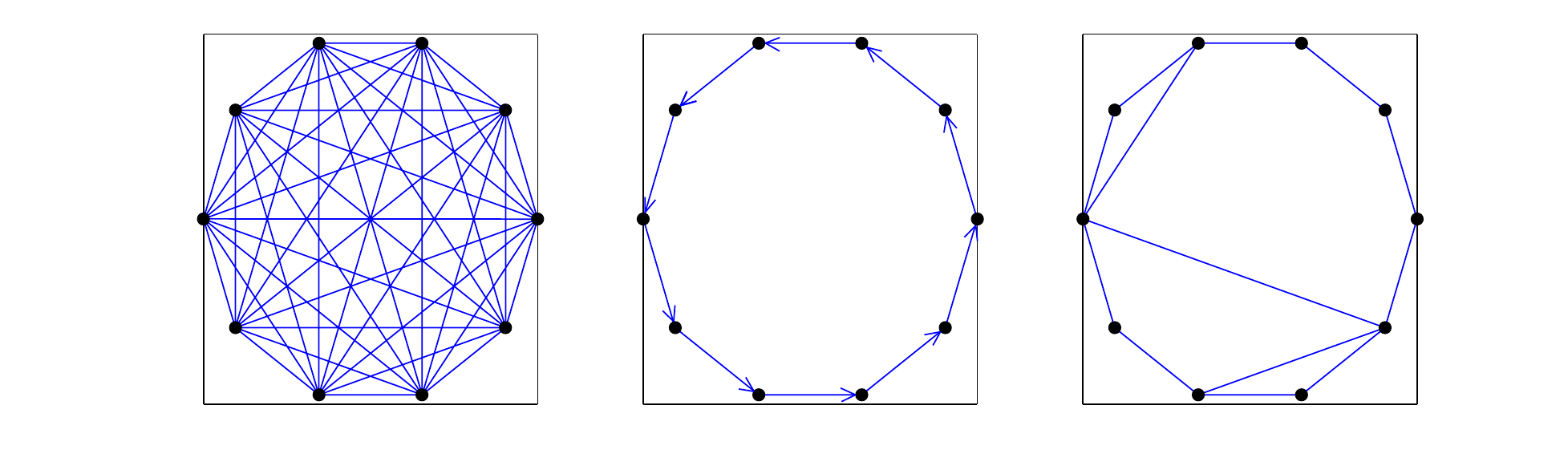}\\
\includegraphics[width=0.4\textwidth]{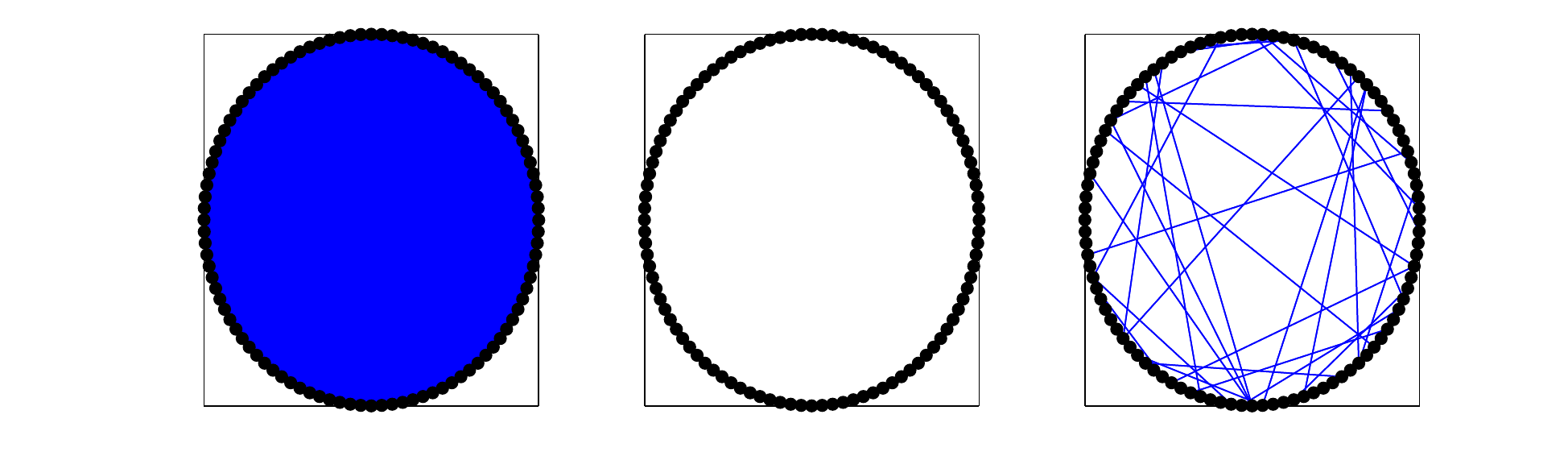}
\end{center}
\caption{Three different network topologies: fully connected, directed ring and undirected small world. The top line is for a population of  $N=10$, the bottom for $N=100$ agents. The small world networks have been generated by adding undirected shortcut links to the undirected ring topology, each with probability $0.3$ \cite{watts1998collective}. The weights have been assigned so that the resulting $P$ matrix is doubly stochastic.  The corresponding matrices verify $\|P\|=1$.  }
\label{fig:od_net}
\end{figure}

\begin{figure}
\begin{center}
\includegraphics[width=0.45\textwidth]{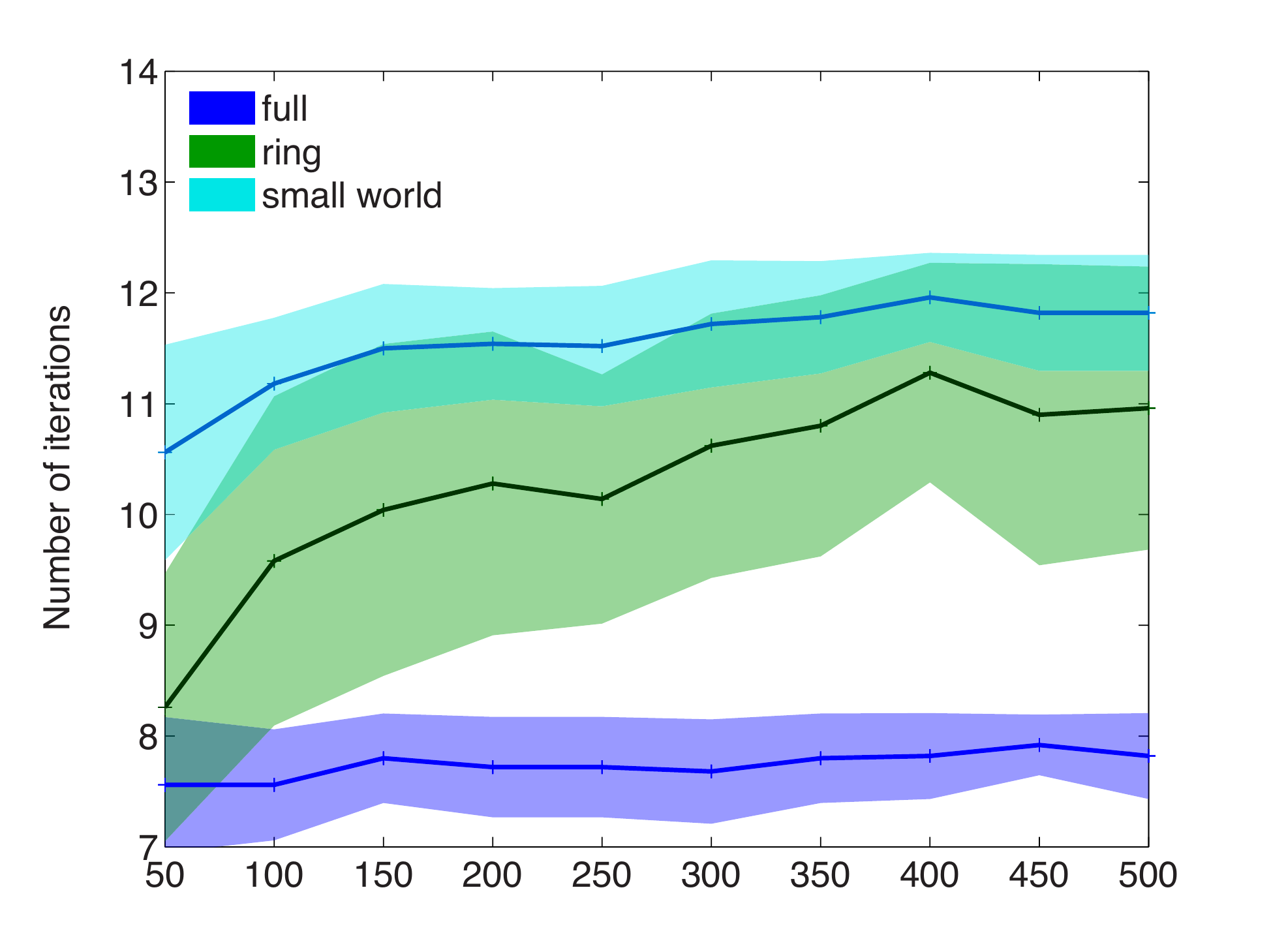}\\
\vspace{-0.5cm}
\includegraphics[width=0.45\textwidth]{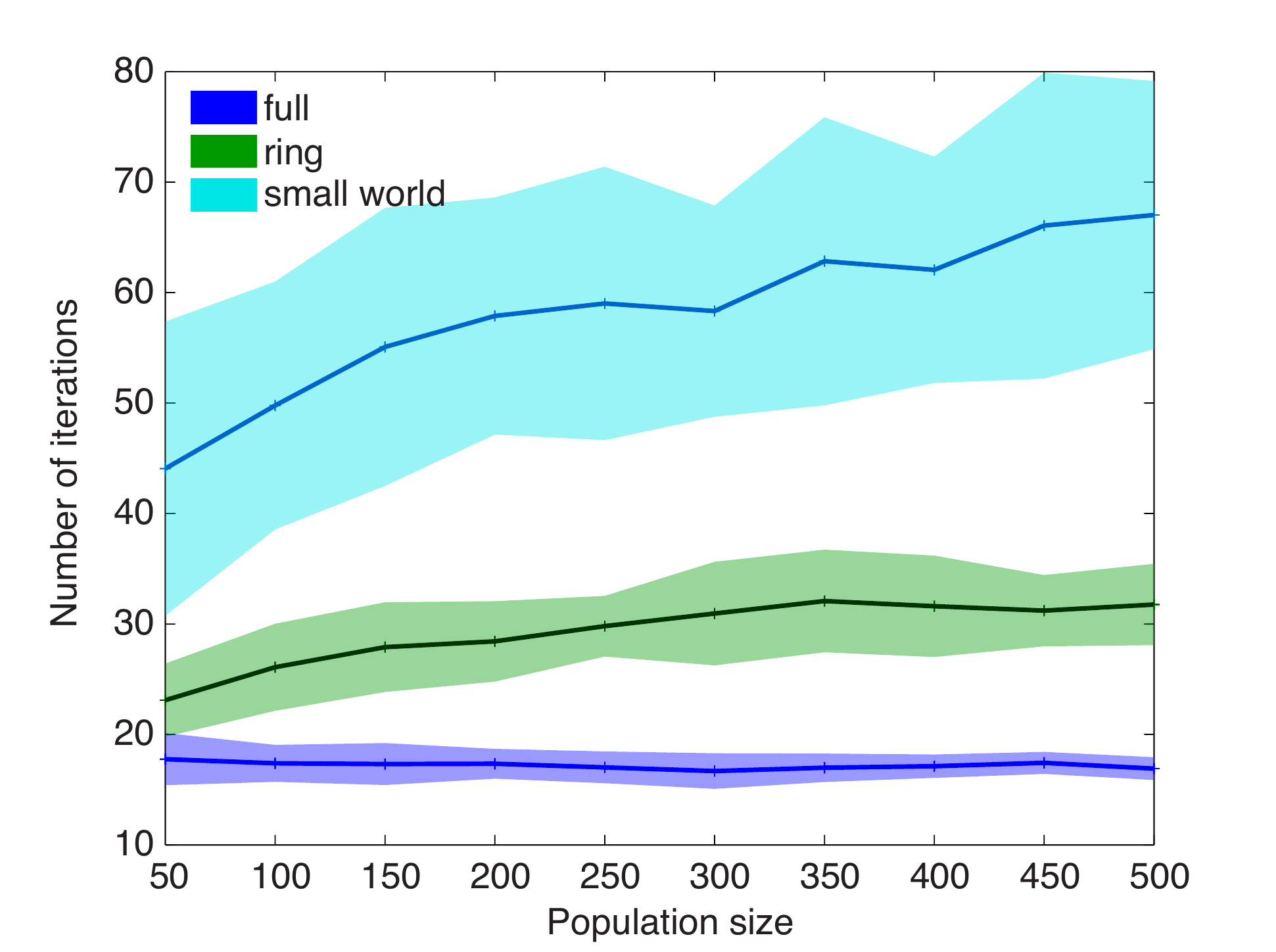}
\caption{ Average number of iterations (solid line) and $90\%$ confidence intervals as a function of the population size $N$ for the three different network topologies of Figure \ref{fig:od_net}. The plot on the top refers to a population where only partially stubborn agents are present and update their opinion using the BR scheme $(\Phi^{\textup{P--B}})$, while the plot on the bottom  refers to a population composed by half  partially stubborn and half follower agents, using the scheme with memory $(\Phi^{\textup{K}})$. In each case, $50$ different networks and populations were simulated with initial opinions chosen according to a uniform distribution in $\left[0,1\right]^2$. For each stubborn agent the  initial opinion was  projected on $\mathcal{X}^i$ to guarantee feasibility of the constraints in~\eqref{opinion}. The stopping criterion is $\|\ze_{(k)}-\ze_{(k-1)}\|_{\infty}\!\le \!10^{-5}$.}
\label{fig:od_pb}
\end{center}
\end{figure}

To investigate the performance of the two schemes we consider a case study where each agent $i$ has $n=2$ opinions $x^i=\left[x^i_1,x^i_2\right]^\top$, regarding two different topics, taking values in $\mathcal{X}^i:=\{\left[x_1, x_2\right]^\top \mid \|x_1-x_2\|^2\le 0.3\}$ and is either a follower or partially stubborn with $\theta_i=1$. Figure~\ref{fig:od_pb} reports the number of iterations required to reach convergence as a function of the population size $N$ for two different compositions of the population and three different network topologies, illustrated in Figure \ref{fig:od_net}. These simulations show that the convergence speed depends only mildly on the population size, making our approach scalable.

\subsection{Demand-response methods and mean field control with local communication} \label{sec:hvac}

By assuming an affine dependence of the cost on the population average, that is, 
\begin{equation} \label{eq:price-function-DR}
p(\bar\sigma_t):=\lambda \bar\sigma_t+p_0, \quad \lambda>0,
\end{equation}
the game in \eqref{eq:demand_response} can be rewritten as the MF game in~\eqref{eq:mfg} with $x^i=u^i$ and cost function as in \eqref{eq:cost} with $\sigma^i=\bar{\sigma}$, $q_i=\rho_i$, $Q= I_T, C=\frac{\lambda}{2}I_T$,  $c=\frac{p_0}{2}-\theta\hat{x}^i$. Note that, since $s(u)$ is an affine function of the input $u$, the constraint $[s(u);u]\in(\mathcal{S}^i\times\mathcal{U}^i )\cap \mathcal{C}^i$ 
 can  be rewritten as a unique  constraint on the input, $u^i=x^i\in\mathcal{X}^i$, that is convex and compact if $\mathcal{U}^i,\mathcal{S}^i,\mathcal{C}^i$ are convex and compact. The previous theory can  be used to find a MF $\varepsilon$-Nash equilibrium using local communications. 
\vspace{0.2cm}
\begin{corollary}\label{cor:demand_response}
Suppose  that Assumptions~\ref{sa},~\ref{a1} and~\ref{a2} hold, and let $p(\bar\sigma_t)$ be as in \eqref{eq:price-function-DR}. The following iterations and conditions guarantee convergence of the strategies computed according to Algorithm~\ref{alg:mem}, from any initial point, to 
a MF $\varepsilon_{N,\nu}$-Nash equilibrium for~\eqref{eq:demand_response}.

\begin{center}\begin{tabular}{llccccr}
&& eq.& $(\nu_1,\nu_2)$ & cost ($\forall i$) & network \\ \cmidrule[.75pt](lr){1-6}
1. & $\Phi^{ \text{P--B} }$ & \eqref{eq:Picard-Banach} &$(0,\nu)$  & $\rho_i>\frac{\lambda}{2}$ & $\|P\|\le1$;& \\
2. & $\Phi^{ \text{K} }$ & \eqref{eq:Krasnoselskij} &$ (0,\nu)$  & $\rho_i\ge\frac{\lambda}{2}$ & $\|P\|\le1$; &\\
3. & $\Phi_k^{ \text{M} }$ & \eqref{eq:Mann} & $(\nu/2,\nu/2)$ & &$P=P^\top$.& $\square$   \\ \cmidrule[.75pt](lr){1-6}
\end{tabular}\end{center}

\end{corollary}
\vspace{0.2cm}

The model given in \eqref{eq:demand_response} can be used to describe demand-response methods for  heating ventilation air conditioning (HVAC) systems in smart buildings, as suggested in~\cite{ma2014distributed}, by selecting $\rho_i=\theta\gamma_i^2$,
where $\theta>0$ is the cost coefficient of the Taguchi loss function and $\gamma_i>0$ specifies the thermal characteristic   of the HVAC system.
In~\cite[Theorems 1, 2]{ma2014distributed} it is shown that, for $N>3$, if $\gamma^i=\gamma>0$ for all $i$, $\mathcal{U}^i=\left[u^i_{\textup{min}}, u^i_{\textup{max}}\right]\subset \R^n$ with $u^i_{\textup{min}}, u^i_{\textup{max}}\in\R^n$, and 
$\lambda\le\frac{ 2\theta\gamma^2}{N-3} $,
then the Nash equilibrium is unique and can be computed using a  control algorithm involving a central coordinator. Corollary~\ref{cor:demand_response} proves that Algorithm~\ref{alg:mem}, on the other hand, can be used to find an $\varepsilon$-Nash equilibrium in a distributed fashion, under less stringent conditions. Firstly, it allows arbitrary convex constraints $u^i\in\mathcal{U}^i$, 
instead of box constraints, hence  including the case of joint stage constraints, ramping and more general dynamics. Secondly, convergence is guaranteed using only local communications over a network that satisfies $\|P\|\le1$ and Assumption~\ref{a2}, instead of requiring the presence of a central coordinator. Thirdly, it demands  the less stringent assumption $\lambda<2\theta\gamma_i^2$ for all $i$ (or no assumption at all  if additionally $P=P^\top$), making the problem scalable for large population sizes.  As a drawback, a MF $\varepsilon$-Nash equilibrium is reached instead of an exact one, with $\varepsilon$  arbitrarily small for large populations. As a particular case we consider a hierarchical communication structure that models the fact that groups of buildings are managed by the same company. For simplicity, let us assume that there are $M$ companies and each one manages $B$ buildings, for a total of $N=MB$ buildings. At every communication step the managers compute the aggregate power demand of their buildings, then communicate among each other using a network $P_M\in\R^{M\times M}$ and finally compute the price signal for their buildings. With the convention that buildings controlled by the same manager are grouped together in the extended vector and that the manager is the first agent of the corresponding block, the above scheme corresponds to the overall weighted adjacency matrix
\begin{align}
P&=\!\!\underbrace{\vphantom{ \left[\begin{smallmatrix}1/B  \\0  \\\tvdots  \\0 \end{smallmatrix}\right] }\left(I_M\otimes \left[\begin{smallmatrix}1 & 0 & \hdots & 0 \\ \tvdots& \tvdots &\tvdots & \tvdots\\1 & 0 & \hdots &  0\end{smallmatrix}\right]\right)}_{\textup{price dispatch}} \underbrace{\vphantom{ \left[\begin{smallmatrix}1/B  \\0  \\\tvdots  \\0 \end{smallmatrix}\right] }\left(P_M\otimes I_B\right)}_{\textup{manager comm.}} \underbrace{\left(I_M\otimes\left[\begin{smallmatrix}1/B & \cdots & 1/B \\0 & \hdots & 0 \\\tvdots &\tvdots & \tvdots \\0 & \hdots & 0\end{smallmatrix}\right]\right)}_{\textup{local aggregation}}\notag\\&=\textstyle P_M\otimes {\frac{1}{B} \mathbbm{1}_{B}\mathbbm{1}_{B}^\top }\label{Ph}.
\end{align}

Lemma~\ref{lem:matrix} in the Appendix, proves that if $P_M$ satisfies the connectivity conditions of Assumption~\ref{a2} then also the matrix $P$ in~\eqref{Ph} does and that $\|P_M\|_2\le1$ implies $\|P\|\le1$. Moreover, if $P_M=P_M^\top$ then $P=P^\top$. It follows from Corollary~\ref{cor:demand_response} that such hierarchical communication structure can be used, instead of a central coordinator, to steer the population to an $\varepsilon$-Nash equilibrium.

\begin{figure}[h!]
\begin{center}
\includegraphics[width=0.45\textwidth]{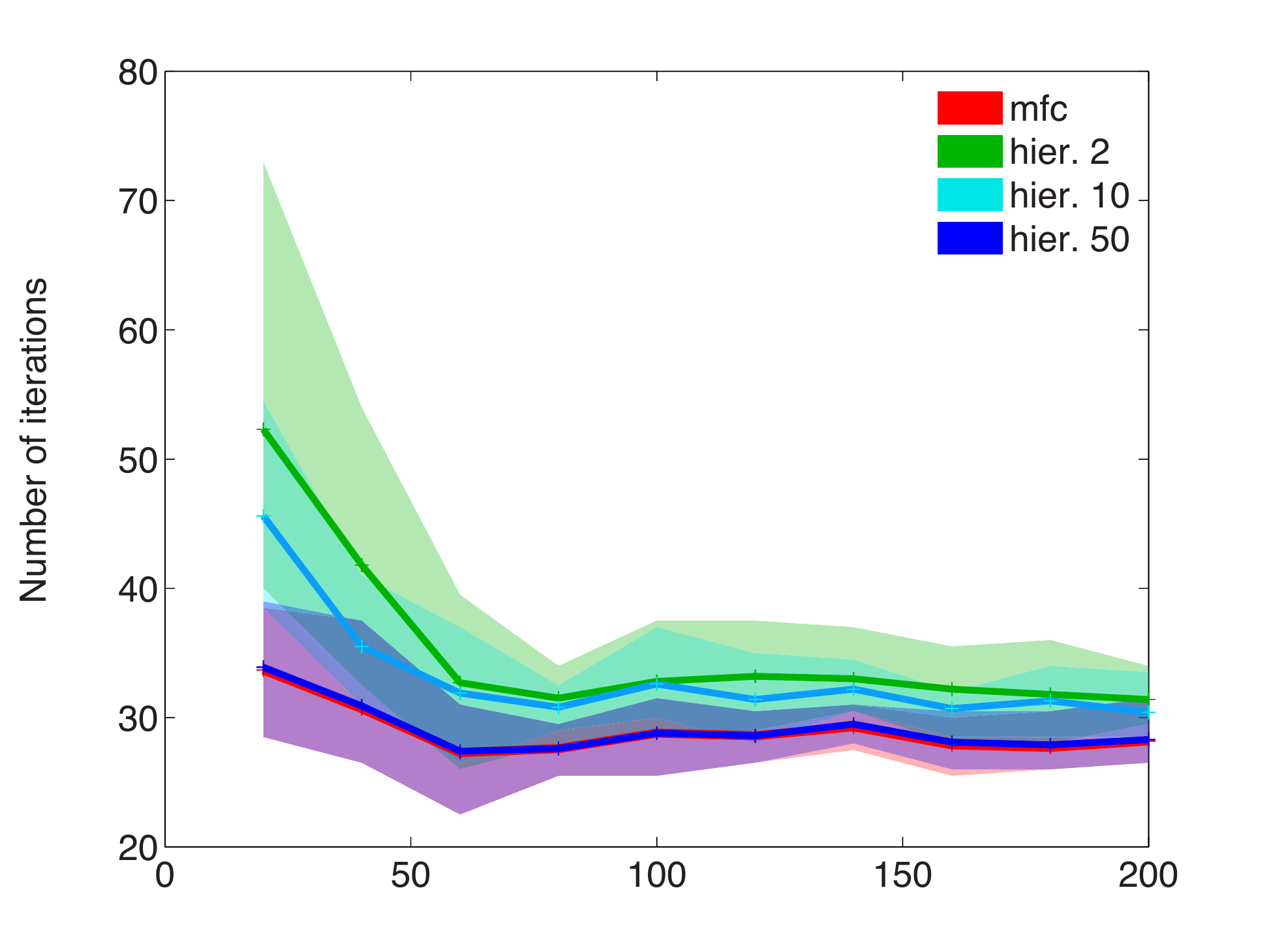}\\
\vspace{-0.5cm}
\includegraphics[width=0.45\textwidth]{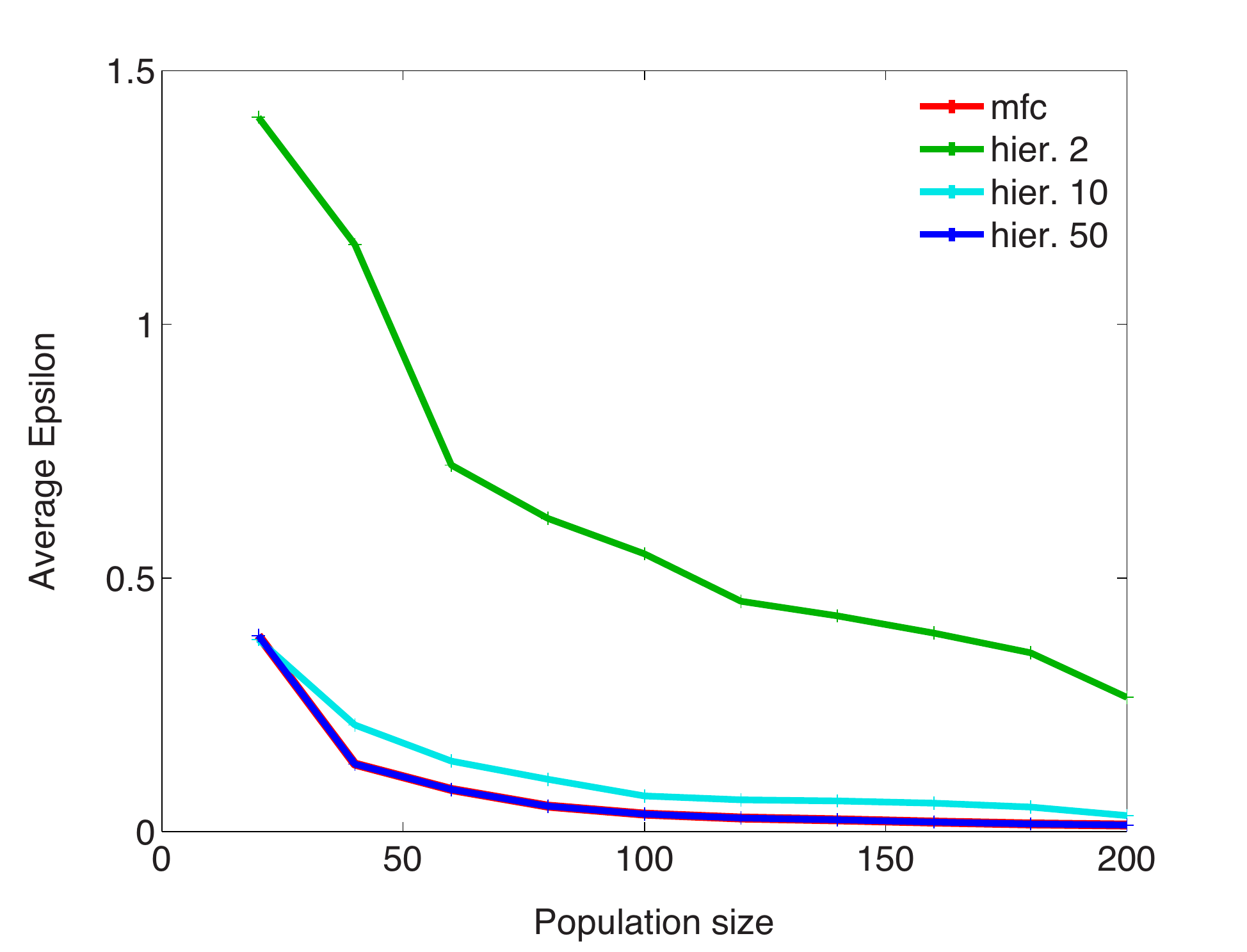}
\caption{ Comparison between the MF control approach with central unit and the hierarchical distributed approach with $M=5$ and $\nu=2,10,50$. The matrix $P_M$ corresponds to a symmetric ring network.
The plot on the top shows the required average number of iterations  needed to reach convergence (and the $90\%$ confidence intervals), 
the one on the bottom the average cost improvement $\varepsilon_N:=\max_{i\in\{1,\ldots,N\}} \bar{J}^i-J^{i\,\star}$, 
where $\bar J^i:=J\left(\bar x^i,\frac1 N \sum_{j=1}^N \bar{x}^j\right)$ and $J^{i\,\star}:=\min_{x^i\in\mathcal{X}^i} J\left(x^i,\frac1N\left(x^i+\sum_{j\neq i}^N \bar{x}^j\right)\right)$, that an agent can achieve by unilateral deviations, both as a function of the population size $N$.
We set  $\lambda=2$ and $\theta \gamma_i^2=0.1$ for all $i$. We consider an horizon of $T=24$ hrs and we assume a baseline energy consumption $\sigma_0$ as illustrated in~\cite[Figure 1]{ma:callaway:hiskens:13}. We set the baseline energy price to $p_0:=\lambda \sigma_0$. The average is computed over $10$ different populations with  prescheduled energy consumption $\hat x^i$ uniformly sampled in  $\left[0,1\right]^T$ and constraint set $\mathcal{X}^i:= \{x\in\R_{\ge0}^T \mid \sum_{t=1}^T x_t^i=\sum_{t=1}^T \hat x^i_t, x^i_t =0 \mbox{ if } t\notin\left[T^i_{\textup{start}}, T^i_{\textup{end}}\right]\}$, with $T^i_{\textup{start}}$ uniformly sampled in $\{1, T\}$, $T^i_{\textup{end}}$ uniformly sampled in $\{T^i_{\textup{start}}+1, T\}$. To guarantee convergence, we use $\Phi_k^{\textup{M}}$ and $\nu_1=\nu_2=\frac{\nu}{2}$. The stopping criterion is
$\|\mn(\m_{\nu_1,0}(\ze_{(k)}))-\m_{\nu_1,0}(\ze_{(k)})\|_{\infty}\le 10^{-3}$.}
\label{fig:b2}
\end{center}
\end{figure}

In Figure \ref{fig:b2} we report the number of iterations required to reach convergence, as a function of the population size $N$, using Algorithm~\ref{alg:mem} with the mapping $\Phi^{\textup{M}}$ and different number of communications, and we compare it with the performance of the traditional centrally coordinated MF control scheme. As in the opinion dynamics case study, the required number of iterations is almost independent on the population size. Moreover, one can notice that already for small values of $\nu$ the solution to the distributed approach performs similarly to the solution obtained with the central coordinator. The higher the number of communications is between two optimization steps the smaller is $\varepsilon$, as guaranteed by  Theorem~\ref{thm:problem 1}, and the fewer iterations of Algorithm~\ref{alg:mem} are required.

We conclude this section by noting that the model given in~\eqref{eq:demand_response}  can be also used, by setting $\rho_i$ sufficiently small, to compute the optimal charging strategy for large populations of Plug-in Electrical Vehicles (PEV)~\cite{ma:callaway:hiskens:13,parise:colombino:grammatico:lygeros:14}. Under the same conditions of~\cite[Corollary 3]{grammatico:parise:colombino:lygeros:14}, Corollary~\ref{cor:demand_response} (point 3.) allows to recover a MF $\varepsilon$-Nash equilibrium using a symmetric network~$P$ instead of a central coordinator.


\section{Conclusion and Outlook} \label{sec:conclusion}

We have considered NA games for populations of agents with different individual behaviors, constraints and interests, and affected by the aggregate behavior of their neighbors. We have characterized the Nash equilibria of such problems and studied the convergence properties of the BR dynamics. For cases where it was not possible to guarantee convergence of the BR dynamics, we have proposed new strategy-update schemes. Finally, we have shown how the NA setting can be extended, by allowing multiple rounds of communications, to provide a distributed solution to the MF control problem.

Our technical results are derived for agents that update their strategies synchronously and over a fixed network. As future work, we believe it would be interesting to study whether similar convergence results can be achieved via asynchronous updates and time-varying or random communications \cite{fagnani2008randomized}, as for example using gossip-based communication  schemes. For the multiple communication case, known results from consensus theory could be  applied to guarantee that the convergence requirement $\sigma^i_{\nu\,(k)}\rightarrow \bar{\sigma}$ is achieved asynchronously. 

The concept of social global optimality \cite{huang:caines:malhame:12} has not been considered in this paper. It would be valuable to characterize the influence of the network structure on the global properties of the associated NA Nash equilibrium. In other words, one may favor networks with given properties in other to coordinate the agents to a desired emergent behavior.

As in traditional aggregative games we have addressed a deterministic setting; a valuable extension would be a stochastic setting where, for instance, the parameters of each agent are extracted from a probability distribution and/or a random input enters in the dynamics, as assumed in classical MF games \cite[Section V]{huang:caines:malhame:07}, \cite[Equation 2.1]{huang:caines:malhame:07}.

Applications of our methods and results include distributed control and game-theoretic coordination in network systems. Among others, application domains that can be further explored  are multi-dimensional opinion dynamics on social networks \cite{etesami2014game, lorenz2007continuous, sergey2015novel,  mirtabatabaei2012opinion}, distributed dynamic demand-side management of aggregated loads in power grids
\cite{ chen2014autonomous, ma2014distributed}, congestion control over networks \cite{barrera:garcia:15}, synchronization and frequency regulation among populations of coupled oscillators \cite{yin:mehta:meyn:shanbhag:12, dorfler:bullo:14},

\section*{Appendix A}
\label{app:operator-theory}

The main idea behind the proof of Theorem~\ref{convergence} is to derive sufficient conditions on the matrices defining the cost function~\eqref{eq:cost} and on the network $P$ to guarantee that the extended aggregation mapping $\mn(\cdot)$ possesses one of the regularity properties listed in the following definition.

\begin{definition}[{Regularity properties}]
\label{def:CON}
Consider the Hilbert space $\mathcal{H}_{S}$ defined by the matrix $S\in\R^{n\times n}$, $S \succ 0$. A mapping $f: \R^n \rightarrow \R^n $ is
\begin{enumerate}
\item  a Contraction  (CON)~\cite[Definition 1.6]{berinde} in $\mathcal{H}_{S}$  if  there exists $\epsilon \in (0,1]$  such that $$\left\| f(x) - f(y) \right\|_S \leq (1-\epsilon) \left\| x-y \right\|_S ,\quad \forall  x, y \in \R^n.$$
\item  Non-Expansive  (NE)~\cite[Definition 4.1 (ii)]{bauschke:combettes} in $\mathcal{H}_{S}$ if $$\left\| f(x) - f(y) \right\|_S \leq  \left\| x-y \right\|_S,\quad \forall   x, y \in \R^n.$$
\item   Firmly Non-Expansive  (FNE)~\cite[Definition 4.1 (i)]{bauschke:combettes} in $\mathcal{H}_{S}$  if for all $x, y \in \R^n$ $$\left\| f(x) - f(y) \right\|^2\! \leq \left\| x-y \right\|_S^2 - \left\|  f(x)\! -\! f(y)\! - \!\left( x\!-\!y\right) \right\|_S^2.$$
\item   Strictly Pseudo-Contractive   (SPC) ~\cite[Remark 4, pp. 13]{berinde} in $\mathcal{H}_{S}$  if there exists $\rho<1$ s.t. for all $x, y \in \R^n$  $$ \left\| f(x) - f(y) \right\|_S^2\! \leq\! \left\| x-y\right\|_S^2  + \rho\! \left\| f(x)\! -\! f(y)\! -\! \left(x\!-\!y \right)\right\|_S^2.$$
\end{enumerate}
\hfill{$\square$}

\end{definition}

Contractiveness is a quite restrictive property, nonetheless this is the property commonly used in the MF control literature, see for example~\cite{huang:caines:malhame:07,  ma:callaway:hiskens:13}. NE mappings are the simplest generalization of CON mappings.
In the technical proofs we will make use of the following equivalent characterization of NE affine mappings.
\vspace{0.2cm}
\begin{lemma} \label{lem:NE}
Consider an affine mapping $f: \R^n \rightarrow \R^n$, $x\mapsto f(x):=Fx+b$, $F\in\R^{n\times n},b\in\R^{n}$ and $R\in\R^{n\times n}, R\succ 0$. The following statements are equivalent:
(1) $f$ is NE in $\mathcal{H}_R$.
(2) $\|F\|_R\le1$.
(3) $F^\top R F-  R \preccurlyeq 0$.
\end{lemma}
\begin{proof}
$(1)    \Leftrightarrow  \|Fr-Fs\|_R\le\|r-s\|_R \ \forall r,s    \Leftrightarrow  \|F(r-s)\|_R\le\|r-s\|_R\  \forall r,s  \Leftrightarrow  \|Fx\|_R\le\|x\|_R\  \forall x  \Leftrightarrow (2)      \Leftrightarrow   \|F x\|_R^2\le\|x\|_R^2\  \forall x  \Leftrightarrow  x^\top F^\top R F x\le x^\top R x \ \forall x  \Leftrightarrow  x^\top (F^\top  R F -R)x\le 0 \ \forall x    \Leftrightarrow  (3)$.
\end{proof}
\vspace{0.2cm}

FNE mappings are a particular subclass of NE mappings, that includes for example the metric projection onto a closed convex set $\text{Proj}_{\mathcal{C}}: \R^n \rightarrow \mathcal{C} \subseteq \R^n$~\cite[Proposition 4.8]{bauschke:combettes}.
In the technical proofs we will make use of the following equivalent characterization of FNE mappings.
\vspace{0.2cm}
\begin{lemma}[{\cite[Lemma 5]{grammatico:parise:colombino:lygeros:14}}] \label{lem:FNE}
A mapping $f: \R^n \rightarrow \R^n$ is FNE in $\mathcal{H}_{S}$ if and only if
$ \left\| f(x) - f(y) \right\|_{S}^{2} \leq \left( x-y\right)^\top S \left( f(x) - f(y)\right), \quad \forall x, y \in \R^n.$
\hfill{$\square$}
\end{lemma}
\vspace{0.2cm}
Finally, SPC mappings are a generalization of NE mappings. An equivalent characterization of SPC mappings can be given in terms of monotone mappings.
\vspace{0.2cm}
\begin{definition}[{Monotonicity}] \label{def:SAC}
A mapping $f: \R^n \rightarrow \R^n$ is 
\begin{enumerate}
\item Strongly monotone (SMON)~\cite[Definition 22.1]{bauschke:combettes} in $\mathcal{H}_S$ if there exists $\epsilon > 0$ such that $\forall x, y \in \R^n,$ $\left( f(x) - f(y) \right)^\top S \left( x-y\right) \geq \epsilon \left\| x-y\right\|_S^{2}.$
\item Monotone (MON)~\cite[Definition 20.1]{bauschke:combettes} in $\mathcal{H}_S$ if
$\quad 
\left( f(x) - f(y)\right)^\top S \left( x-y\right) \geq 0,\quad \forall x, y \in \R^n.
${\hfill $\square$}
\end{enumerate}

\end{definition}

\vspace{0.2cm}
\begin{lemma}[{\cite[Lemma 1 and 2]{grammatico:parise:colombino:lygeros:14}}] \label{lem:SA} 
1) If $f: \R^n \rightarrow \R^n$ is MON  and $g: \R^n \rightarrow \R^n$ is SMON in $\mathcal{H}_S$, then $f+g$ is SMON in $\mathcal{H}_S$.
2) If $\text{Id} - f$  is Lipschitz and SMON in $\mathcal{H}_S$, then $f$ is SPC in $\mathcal{H}_S$.
\hfill{$\square$}
\end{lemma}

\section*{Appendix B}
\label{app:proofs}


\subsection*{Proof of Theorem~\ref{thm:problem 2} (Solution to Problem 1)}
Since $x^{i\,\star}(z^i)=\text{Proj}_{\mathcal{X}^i}^{q_iQ}(-(q_iQ)^{-1}(Cz^i+c_i))$ in~\eqref{eq:opt_prob} is a continuous mapping in $z^i$ the mapping $\xe(\ze)$  in~\eqref{eq:x_extended} is continuous in $\ze$. Consequently, the mapping $\mn(\ze)= \Pen \xe(\ze) $  is continuous. 
By Assumption~\ref{sa}, the set $\boldsymbol{\mathcal{P}}_\nu\mathcal{X}_{1\times N}$ is compact and convex. Therefore, by the Brouwer fixed point theorem~\cite[Theorem 4.1.5]{smart1980fixed} the mapping $\mn$ admits at least one fixed point. 
Consider now an arbitrary fixed point $\bar \ze=\left[\bar z^1;\hdots;\bar z^N\right]\in\R^{Nn}$ of the aggregation mapping $\mn$ in~\eqref{eq:mn}, that is $\bar z^i = \sum_{j=1}^{N} P^{\nu}_{ij}x^{j \, \star}(\bar{z}^j)$. According to Definition~\ref{def:Nash-equ}, the set of strategies $\left\{ \bar{x}^i := x^{i \, \star}( \bar{z}^i ) \right\}_{i=1}^{N}$ is a NA Nash equilibrium  if 
 any agent $i$, given the strategies $\left\{ \bar{x}^j  \right\}_{j\neq i}^{N}$ of all the other agents, cannot improve its cost, that is
$\bar{x}^{i} =\argmin_{y \in \mathcal{X}^i } J^i\left( y, P^{\nu}_{ii}y+\sum_{j\neq i}^N P^{\nu}_{ij}\bar x^{j}\right).$
By definition of fixed point and using the fact that $P^{\nu}_{ii}=0\, \forall i$
\begin{align*}\bar{x}^{i} &= x^{i \, \star}( \bar{z}^i )\! = \!\argmin_{y \in \mathcal{X}^i } J^i ( y,\bar z^{i})\! =\!\!  \argmin_{y \in \mathcal{X}^i } J^i( y, \textstyle\sum_{j=1}^{N} \! P^{\nu}_{ij} \bar x^{j}  )\\
&\textstyle =\! \argmin_{y \in \mathcal{X}^i } J^i( y, P^{\nu}_{ii}\bar x^{i}\!+\!\sum_{j\neq i}^N \!P^{\nu}_{ij}\bar x^{j}) \!
\\&\textstyle=\!\argmin_{y \in \mathcal{X}^i } \textstyle  J^i( y, P^{\nu}_{ii}y+\sum_{j\neq i}^NP^{\nu}_{ij}\bar x^{j}).\hspace{2.1cm} \blacksquare
\end{align*}


\subsection*{Proof of Lemma~\ref{lemma:SA} (Properties of the extended aggregation mappings)}
The fact that $\m(\ze)= \Ie\, \xe(\ze)$ admits at least one fixed point can be proven as done for $\mn(\ze)= \Pen \xe(\ze) $ in the proof of Theorem~\ref{thm:problem 2}.
 Let $ D_N:=\max_{\boldsymbol{x}\in\mathcal{X}_{1\times N}} \|\boldsymbol{x}\|$; then for any $\ze\in\R^{Nn}$
$\|\mn(\ze)-\m(\ze) \|=\|\Pen \xe(\ze)-\Ie \xe(\ze) \|\le \|\Pen-\Ie\|\|\xe(\ze)\|\le \|\Pen-\Ie\| D_N.$
Hence 
$\sup_{\ze\in\R^{Nn}}      \|\mn(\ze)-\m(\ze) \|\le \|\Pen-\Ie\| D_N. $
By Assumption~\ref{a2}, $\lim_{\nu\rightarrow\infty}\|P^\nu-\ones\|=0$ which implies that $\lim_{\nu\rightarrow\infty} \|\Pen-\Ie\|=\lim_{\nu\rightarrow\infty} \|(P^\nu-\ones)\otimes I_n\|=0$, therefore 
$\lim_{\nu\rightarrow\infty} \sup_{\ze\in\R^{Nn}}        \|\mn(\ze)-\m(\ze)\|=0.$ \hfill $\blacksquare$


\subsection*{Proof of Lemma~\ref{th:continuityFixedPoints} (Fixed point sets)}
The proof of this lemma follows the lines of~\cite[Lemmas 7.42, 7.44]{kirk2001handbook}.
In particular, we start from the following lemma.
\begin{lemma}
Under Assumptions~\ref{sa} and~\ref{a2}, given two sequences $(\nu_h \in\N)_{h=1}^\infty$, $(\bar \ze_{h}\in\R^{Nn})_{h=1}^\infty$ such that  $\lim_{h\rightarrow\infty}\nu_h=\infty$,
 $\bar \ze_{h}=\mn_{_{h}}(\bar \ze_{h}), \forall h\ge1$ and
$\lim_{h\rightarrow\infty} \bar \ze_{h}=\bar \ze $, then $\bar \ze=\m(\bar \ze)$. 
\label{lemma:xfp}
\end{lemma}
\begin{proof}
For the sake of contradiction, assume  $\bar \ze\neq\m(\bar \ze)$. Then there exists $\varepsilon>0$ such that
$\mathcal{B}(\bar \ze,\varepsilon)\cap \mathcal{B}(\m (\bar \ze),\varepsilon)=\varnothing,$
where $\mathcal{B}(\bar \ze,\varepsilon):=\left\{\bold{x} \in\R^{Nn}\mid \|\bold{x}-\bar \ze\|\le\varepsilon\right\}$.
We know that the following statements hold:
1) Since, by Lemma~\ref{lemma:SA}, $\mn(\cdot)$ converges uniformly to $\m (\cdot)$, 
$\exists H_1>0: \ \m _{\nu_h}(\bold{u})\in \mathcal{B}(\m(\bold{u}),\varepsilon/2) \quad \forall h\ge H_1, \forall \bold{u}\in\R^{Nn};$
2) Since $\m (\cdot)$  is continuous and $\bar \ze_{h}\rightarrow \bar \ze$,
$\exists H_2>H_1: \ \m (\bar \ze_{h})\in \mathcal{B}(\m (\bar \ze),\varepsilon/2) \quad \forall h\ge H_2;$
3) Since  $\bar \ze_{h}\rightarrow \bar \ze$, 
$\exists H_3>H_2: \ \bar \ze_{h} \in \mathcal{B}(\bar \ze,\varepsilon) \quad \forall h\ge H_3.$
Therefore, from point 1) with $\boldsymbol{u}=\bar{\boldsymbol{z}}_h$ and 2) we get that, for all $h\ge H_3$,
$\bar \ze_{h}=\m _{\nu_h}(\bar \ze_{h})\in \mathcal{B}(\m (\bar \ze_{h}),\varepsilon/2)\subseteq \mathcal{B}(\m (\bar \ze),\varepsilon).$
Hence, using 3), we get
$\bar \ze_{h}\in \left( \mathcal{B}(\m (\bar \ze),\varepsilon)\cap \mathcal{B}(\bar \ze,\varepsilon) \right)\neq \varnothing,$ for all $h\ge H_3$,
which is a contradiction.
\end{proof}
\vspace{0.2cm}
We  now prove Lemma~\ref{th:continuityFixedPoints}.
For the sake of contradiction, suppose that the claim is not true.
Then there exists $ \epsilon>0$ and a sequence $\left(\nu_h\right)_{h=1}^\infty$, with $\nu_h\rightarrow \infty$, such that $\forall h\ge1$ $\mu(\mathcal{F}_h,\mathcal{F})> \epsilon$. Equivalently, $\forall h\ge1\ \exists \bar \ze_{h}\in \mathcal{F}_{\nu_h}$  such that $\mu(\bar \ze_{h},\mathcal{F})> \epsilon$. Note that $\bar{\boldsymbol{z}}_h=\mn_{_h}(\bar{\boldsymbol{z}}_h)\in \Pen_{_h}\mathcal{X}_{1\times N} \subset \left(\textup{conv}\{\mathcal{X}^i\}_{i=1}^N\right)^N=:\mathcal{X}^N$ for all $h\in\N$, and $\mathcal{X}^N$ is compact by Assumption~\ref{sa}. Since any sequence defined in a compact set has a convergent subsequence, we can assume without loss of generality, that $\lim_{h\rightarrow\infty} \bar \ze_{h}=\bar \ze\in{\mathcal{X}^N}$. Note that 
$\mu(\bar \ze_{h},\mathcal{F})> \epsilon$ for all $h\ge1$ implies that $\mu(\bar \ze,\mathcal{F})\ge\epsilon$ and hence $\bar \ze \not\in \mathcal{F}$. On the other hand, the sequences  $\left(\nu_h\right)_{h=1}^\infty$ and  $\left(\bar \ze_{h}\right)_{h=1}^\infty$ satisfy the conditions in Lemma~\ref{lemma:xfp}, therefore $\bar \ze=\m(\bar \ze) \Rightarrow \bar \ze \in \mathcal{F}$, which is a contradiction. \hfill 
$\blacksquare$


\subsection*{Proof of Theorem~\ref{thm:problem 1} (Solution to Problem 2)}
We follow a similar argument as in~\cite[Proof of Theorem~1]{grammatico:parise:colombino:lygeros:14}.
Given the fact that $q_i\ge \underline q$ for all $i$, the mappings $x^{i\,\star}(z^i)=\text{Proj}_{\mathcal{X}^i}^{q_iQ}(-(q_iQ)^{-1}(Cz^i+c_i))$ are uniformly Lipschitz with some constant $L_\text{x}>0$~, that is $\|x^{i\,\star}(z_a)-x^{i\,\star}(z_b)\|\le L_\text{x}\|z_a-z_b\|$ for all $i$ and all $z_a,z_b\in\mathcal{X}$. 
Moreover, given the fact that $q_i\le \bar q$ for all $i$, the quadratic cost functions $J^i$ are uniformly Lipschitz in the compact set $\mathcal{X}\times \mathcal{X}$ with some constant $L_J>0$, that is $|J^i(x_a,z_a)-J^i(x_b,z_b)|\le L_J (\|x_a-x_b\|+\|z_a-z_b\|)$ for all $i$ and for all $x_a,x_b,z_a,z_b\in\mathcal{X}$. Let $D=\max_{x\in \mathcal{X}}\|x\|$. For any $(N,\nu)\in\N^2$, consider 
an arbitrary fixed point $\bar \ze=\left[\bar z^1;\hdots;\bar z^N\right]\in\R^{Nn}$ of the aggregation mapping $\mn$ in~\eqref{eq:mn}, that is $\bar z^i = \sum_{j=1}^{N} P^{\nu}_{ij}x^{j \, \star}(\bar{z}^j) $ and define the set of strategies $ \bar x^{i}:=x^{i \, \star}(\bar{z}^i)$. Let $\tilde{x}^{i \, \star}$ denote the optimal strategy, according to problem~\eqref{eq:mfg}, for agent $i$ given that the strategies of the others are fixed to $\left\{ \bar{x}^j \right\}_{j\neq i}^{N}$, that is, 
$\tilde{x}^{i \, \star} := \argmin_{y \in \mathcal{X}^i }\textstyle J^i\left( y, \frac1Ny +  \sum_{j \neq i}^{N}  \frac1N \bar{x}^j  \right)$ and let $
 \tilde{\tilde{x}}^{i \, \star} := \argmin_{y \in \mathcal{X}^i } \textstyle J^i\left( y, \frac1N \tilde{x}^{i \, \star} + \sum_{j \neq i}^{N} \frac1N \bar{x}^j  \right).$
Let us also define the associated costs
$ \bar{J}^{i \, \star} =
\textstyle J^i\left( \bar x^{i }, \frac1N \bar{x}^{i} + \sum_{j \neq i}^{N} \frac1N \bar{x}^j  \right)$, 
$\tilde{J}^{ i \, \star} =  
\textstyle J^i\left( \tilde x^{i \, \star }, \frac1N \tilde x^{i \, \star } + \sum_{j \neq i}^{N} \frac1N\bar{x}^j \right) = \textstyle
\min_{y \in \mathcal{X}^i } J^i( y, \frac1N y + \sum_{j \neq i}^{N} \frac1N \bar{x}^j  )$, 
and $\textstyle \tilde{\tilde{J}}^{ i \, \star} =
\textstyle J^i\left( \tilde{\tilde{x}}^{ i \, \star }, \frac1N \tilde{x}^{i \, \star} + \sum_{j \neq i}^{N} \frac1N \bar{x}^j  \right) = \textstyle
\min_{y \in \mathcal{X}^i } J^i\left( y, \frac1N \tilde{x}^{i \, \star} + \sum_{j \neq i}^{N} \frac1N \bar{x}^j  \right).$ 
Note that $\tilde{\tilde{J}}^{i \, \star} \leq \tilde{J}^{ i \, \star} \leq\bar{J}^{ i \, \star}$.
We define $\tilde{z}^i := \frac1N \tilde{x}^{i \, \star} + \sum_{j \neq i}^{N} \frac1N \bar{x}^j $, so that $\tilde{\tilde{x}}^{i}=x^{i \, \star}( \tilde{z}^i ) $,  and note that 
\begin{align*}
&|\bar{J}^{ i \, \star}-\tilde{J}^{i \, \star}|\le|\bar{J}^{ i \, \star}-\tilde{\tilde{J}}^{i \, \star}|\\&= \textstyle |J^i( \bar x^{i },\frac1N \bar{x}^{i} + \sum_{j \neq i}^{N} \frac1N \bar{x}^j)-J^i(\tilde{\tilde{x}}^{i} ,\frac1N \tilde{x}^{i \, \star} + \sum_{j \neq i}^{N} \frac1N \bar{x}^j )| \\&
\textstyle \le  L_J(\|\bar x^{i }-\tilde{\tilde{x}}^{i}\|+\frac1N\|\bar x^{i }- \tilde{x}^{i \, \star} \|)\\&=L_J(\|x^{i \, \star}(\bar{z}^i) -x^{i \, \star}(\tilde{z}^i) \|+\frac1N\|\bar x^{i }- \tilde{x}^{i \, \star} \|) \\&
 \textstyle\le L_J(L_x\|\bar{z}^i - \tilde{z}^i \|\!+\!\frac1N\|\bar x^{i }- \tilde{x}^{i \, \star} \|) \le L_J(L_x\|\bar{z}^i - \tilde{z}^i \|\!+\!\frac{2D}{N} )
 \end{align*}
\begin{align*}
&\mbox{and} \ \|\bar{z}^i - \tilde{z}^i \|
 \textstyle= \| P_{ii}^\nu\bar x^i - \frac1N \tilde{x}^{i \, \star} + \sum_{j \neq i}^{N}(P_{ij}^\nu- \frac1N) \bar{x}^j  \| \\&
 \textstyle\le \| P_{ii}^\nu\bar x^i - \frac1N \tilde{x}^{i \, \star}\| +\| \sum_{j \neq i}^{N}(P_{ij}^\nu- \frac1N) \bar{x}^j  \|\\
& \textstyle\le \| P_{ii}^\nu\bar x^i -  \frac1N\bar x^i + \frac1N\bar x^i - \frac1N \tilde{x}^{i \, \star}\| + \sum_{j \neq i}^{N}\|(P_{ij}^\nu- \frac1N) \bar{x}^j  \| \\
& \textstyle\le  \sum_{j =1}^{N}\|(P_{ij}^\nu- \frac1N) \bar{x}^j  \|+\frac1N\|\bar x^{i }- \tilde{x}^{i \, \star} \| \\&
 \textstyle\le  D(\sum_{j =1}^{N}|P_{ij}^\nu- \frac1N |+\frac{2}{N}) 
 \textstyle\le D( \| P^\nu-\ones\|_\infty+\frac{2}{N}). 
\end{align*}

Hence 
$|\bar{J}^{ i \, \star}-\tilde{J}^{i \, \star}|\le 2L_J(L_x+1)D\textstyle ( \| P^\nu-\ones\|_\infty+\frac{1}{N} ):=K( \| P^\nu-\ones\|_\infty+\frac{1}{N} ). $
Consequently, an arbitrary agent $i$ can improve its cost at most by an amount $\varepsilon_{N,\nu}:=K(\| P^\nu-\ones\|_\infty   +\frac1N) $ if all other strategies $\left\{ \bar{x}^j := x^{j \, \star}( \bar{z}^j ) \right\}_{j\neq i}^{N}$ are fixed. Therefore, the set of strategies $\left\{\bar{x}^i\right\}_{i=1}^N$ is an $\varepsilon_{N,\nu}$-Nash equilibrium for the MF game in~\eqref{eq:mfg}. $K$ is a constant that does not depend on $N,P$ or $\nu$ and for any fixed $N$ we have $\| P^\nu-\ones\|_\infty \rightarrow 0$ as $\nu\rightarrow \infty$. Consequently,  for all $\varepsilon>0$ and for any fixed $N>\bar N:=\frac{K}{\varepsilon}$, there exists $\bar \nu$ such that  for all $\nu\ge \bar \nu$, we have $\varepsilon_{N,\nu}<\varepsilon$ .\hfill
$\blacksquare$


\subsection*{Proof of Corollary~\ref{cor:problem 1 rate} }
 By the proof of Theorem~\ref{thm:problem 1} the set of strategies $\left\{x^{i\,\star}(\bar{z}^i)\right\}_{i=1}^N$ is an $\varepsilon_{N,\nu}$-Nash equilibrium for the MF game in~\eqref{eq:mfg}, with $\varepsilon_{N,\nu}= K(\frac1N+\| P^\nu-\ones\|_\infty )$.  By the properties of the matrix norm $\| P^\nu-\ones\|_\infty\le \sqrt{N}\| P^\nu-\ones\|_2 =\sqrt{N}\sigma_{\textup{max}}( P^\nu-\ones)=\sqrt{N}\max_{\lambda\in \Lambda( P^\nu-\ones)}|\lambda |=\sqrt{N}\mu_N^\nu\le\sqrt{N}\mu^\nu$, where we used the fact that the matrix $P^\nu-\ones$ is symmetric and,  since $P$ is symmetric, primitive and doubly stochastic, $1$ is a simple eigenvalue and it holds $\Lambda(P^\nu-\ones)=((\Lambda(P))^\nu\backslash\{1\})\cup\{0\}$. Hence
 $\varepsilon_{N,\nu}\le K(\frac1N+\sqrt{N}\mu^\nu )$.


\subsection*{Proof of Theorem~\ref{convergence} (Convergence of Algorithm~\ref{alg:br} and ~\ref{alg:mem})}
To study the convergence of Algorithm~\ref{alg:br} and ~\ref{alg:mem}, we start by studying the regularity properties of the mapping $\mn$.
First, we analyze the regularity properties of the constrained minimizer $x^{i\,\star}$ of problem~\eqref{eq:opt_prob}.

\begin{lemma}[Regularity of the constrained minimizer]\label{lemma:RegularityOptimizer}
 For any $i \in \Z[1,N]$, let $x^{i\,\star}(z^i)$ be as in~\eqref{eq:opt_prob} and $M_i$ as in \eqref{eq:con}. The following facts hold:
1)  If~$M_i\ \succ 0$, then the mapping $x^{i\,\star}$  is a CON in $\mathcal{H}_{Q}$;
2)  If~$M_i \succcurlyeq 0$, then the mapping $x^{i\,\star}$  is NE in $\mathcal{H}_{Q}$;
3) If $-q_iQ \preccurlyeq \C \prec 0$, then the mapping $x^{i\,\star}$  is  FNE in $\mathcal{H}_{-C}$  ;
4) If $0 \prec \C$, then the mapping $-x^{i\,\star}$  is  MON in $\mathcal{H}_{ C}$.
\hfill $\square$
\end{lemma}
\begin{proof}
The proof is an extension to the proof of~\cite[Theorem 2]{grammatico:parise:colombino:lygeros:14} for the case of heterogeneous cost functions and follows from the proof of \cite[Theorem 2]{grammatico2015cdc}.
It is  reported here for completeness.
Note that $x^{i\,\star}(z^i):=\text{Proj}_{\mathcal{X}^i}^{q_iQ}(-(q_iQ)^{-1}(Cz^i+c_i))$.
1) and 2) follow from~\cite[Theorem 2]{grammatico:parise:colombino:lygeros:14} by setting  $Q=q_iQ, \Delta=0$ and by noting that $x^{i\,\star}$ is CON/NE in $\mathcal{H}_{Q}$ if and only if the same holds in $\mathcal{H}_{q_iQ}$.
3) From \cite[Proposition 4.8]{bauschke:combettes} we have that $\text{Proj}_{ \mathcal{X}^i}^{ Q }$ is FNE in $\mathcal{H}_{Q}$, hence by Lemma~\ref{lem:FNE}  in Appendix A we have that, for all $v,w \in \R^n$,
\begin{align}\label{eq:FNE-distorted-projection}
&(  \text{Proj}_{ \mathcal{X}^i}^{ Q}( -(q_iQ)^{-1}\! ( C v\! +\! c_i ) ) -  \text{Proj}_{ \mathcal{X}^i}^{ Q}( -(q_iQ)^{-1}\! ( C w \!+ \!c_i ) ) )^{\top} \notag \\ &\hspace{4.6cm}
\cdot Q \left( -(q_iQ)^{-1} C (v-w) \right)\geq \notag  \\ &
 \! \left\| \text{Proj}_{ \mathcal{X}^i}^{ Q}( -(q_iQ)^{-1}\! ( C v \!+\! c_i ) ) \!-\! \text{Proj}_{ \mathcal{X}^i}^{ Q}( -(q_iQ)^{-1} \!( C w\! +\! c_i ) ) \right\|_{Q}^2 \notag  \\ &\!\Leftrightarrow\! \left( x^{i \, \star}(v)\! - \!x^{i \, \star}(w) \right)\!\!^\top \!
(-q_i^{-1}C) (v-w)\notag\! \!\geq \! \left\| x^{i \, \star}(v)\! - \!x^{i \, \star}(w) \right\|_{Q}^2 
 \notag   \\
  &\!\Leftrightarrow\! \left( x^{i \, \star}(v)\! -\! x^{i \, \star}(w) \right)\!\!^\top \! (-C) (v-w)\! \geq\! \left\| x^{i \, \star}(v) - x^{i \, \star}(w) \right\|_{q_iQ}^2
\end{align}
Therefore, if  $q_i Q \succcurlyeq -C \succ 0$ then from \eqref{eq:FNE-distorted-projection} we get that
$\left( x^{i \, \star}(v) - x^{i \, \star}(w) \right)^\top \! (-C) (v-w) \geq 
\left\| x^{i \, \star}(v) - x^{i \, \star}(w) \right\|_{-C}^2$
which implies by Lemma \ref{lem:FNE} that $x^{i \, \star}$ is FNE in $\mc{H}_{-C}$.
4) From the last inequality in \eqref{eq:FNE-distorted-projection}, we get that
$  \left( -x^{i \, \star}(v) + x^{i \, \star}(w) \right)^\top 
C(v-w)  \geq 0 $
for all $v,w \in\R^n$. If $C \succ 0$, this implies that $-x^{i \, \star}(\cdot)$ is MON in $\mathcal{H}_{C}$.
\end{proof}

Next, we show that the regularity properties of the constrained minimizer $x^{i\,\star}$ in~\eqref{eq:opt_prob} are inherited by the extended mapping $\xe$ in~\eqref{eq:x_extended}.

\begin{lemma}[Regularity of the extended optimizer]\label{lemma:RegularityBigOptimizer}
Let $\xe(\cdot)$ be the extended mapping defined in~\eqref{eq:x_extended}. Given $M_i$ as in \eqref{eq:con}, the following facts hold.
1)  If~$M_i\ \succ 0\ \forall i\in\Z[1,N]$, then the mapping $\xe$  is a CON in $\mathcal{H}_{I_N\otimes Q}$;
2)  If~$M_i \succcurlyeq 0\ \forall i\in\Z[1,N]$, then the mapping $\xe$  is NE in $\mathcal{H}_{I_N\otimes Q}$;
3) If $-q_i Q \preccurlyeq \C \prec 0\ \forall i\in\Z[1,N]$, then the mapping $\xe$  is  FNE in $\mathcal{H}_{I_N\otimes (-C)}$;
4) If $0 \prec \C$, then the mapping $-\xe(\cdot)$  is  MON in $\mathcal{H}_{I_N\otimes C}$.
\hfill $\square$
\end{lemma}
\begin{proof}
\begin{enumerate}
\item For all $i\in\Z[1,N]$, if~$M_i\ \succ 0$ then, 
by Lemma \ref{lemma:RegularityOptimizer},
the mapping ${x}^{i\,\star}$ is a CON in $\mathcal{H}_{Q}$, with some  rate $\delta_i\in\left(0,1\right]$. Therefore, for any $\ae,\be \in \mathbb{R}^{Nn}$ we have
$\|\xe(\ae)-\xe(\be)\|_{I_N\otimes Q}^2 =
\|\left[x^{1\,\star}(\a^1)\!-\!x^{1\,\star}(\b^1); \hdots ; x^{N\,\star}(\a^N)\!-\!x^{N\,\star}(\b^N)\right]\|_{I_N\otimes Q}^2
\\ = \|x^{1\,\star}(\a^1)-x^{1\,\star}(\b^1) \|_Q^2+\hdots+\|x^{N\,\star}(\a^N)-x^{N\,\star}(\b^N) \|_Q^2
\le(1-\delta_1)^2\|\a^1-\b^1\|_Q^2+\hdots+(1-\delta_N)^2\|\a^N-\b^N\|_Q^2\le(1- \min_{i \in \Z[1,N]} \delta_i)^2\|\ae-\be\|_{I_N\otimes Q}^2.
$
Note that $\delta:=\min_{i \in \Z[1,N]}{\delta_i}$ is strictly positive since $N$ is finite.
\item  If~$M_i \succcurlyeq 0$ then, 
by Lemma \ref{lemma:RegularityOptimizer},
the mapping ${x}^{i\,\star}$ is NE in $\mathcal{H}_{Q}$. The proof is the same as in the previous point, with $\delta_i=0$ for all $i$.

\item If $-q_iQ \preccurlyeq \C \prec 0\,\forall i\in\Z[1,N]$ then, 
by Lemma~\ref{lemma:RegularityOptimizer},
the mappings ${x}^{i\,\star}$ are  FNE in $\mathcal{H}_{(-C)}$. Therefore, by Lemma~\ref{lem:FNE},   for all $\ae,\be \in \mathbb{R}^{Nn}$ we have
$
\|\xe(\ae)-\xe(\be)\|_{I_N\otimes (-C)}^2=\sum_{i=1}^N\|x^{i\,\star}(\a^i)-x^{i\,\star}(\b^i) \|_{(-C)}^2\\
\le\sum_{i=1}^N(\a^i-\b^i)^\top(-C)(x^{i\,\star}(\a^i)-x^{i\,\star}(\b^i))\\=(\ae-\be)^\top(I_N\otimes (-C))(\xe(\ae)-\xe(\be)).
$

\item If $0 \prec \C$ then, 
by Lemma \ref{lemma:RegularityOptimizer},
the mappings $-x^{i\,\star}$  are  MON in $\mathcal{H}_{C}$. Therefore,  for any $\ae,\be \in \mathbb{R}^{Nn}$ we have
$
\left( -\xe(\ae)+\xe(\be)\right)^\top(I_N\otimes C)\left( \ae-\be\right)=\sum_{i=1}^N\left( -x^{i\,\star}(\a^i)+x^{i\,\star}(\b^i)\right)^\top C \left(\a^i-\b^i\right)\ge0. $
\end{enumerate}
\end{proof}
The following lemma proves that the linear mapping $\boldsymbol{x}\mapsto \Pen\boldsymbol{x}$ is NE, if $\|P\|\le 1$.

\begin{lemma}\label{lemma:NE_is_preserved}
Consider $P \in \mathbb{R}^{N \times N}, \nu\in\N$ and $\Pen = P^\nu \otimes I_n$. For any $S \in \mathbb{R}^{n \times n} \,, S \succ 0$, let $\Sw := I_N \otimes S$. If $\|P\|\le1$, then  $\|\Pen\|_{\Sw}\le 1$. \hfill $\square$
\end{lemma}

\begin{proof}
The condition $\|P\|\le 1$ implies $\|P^\nu\|\le\|P\|^\nu\le 1$. By Lemma~\ref{lem:NE} ($R=I_N$) this implies $\left( {P^\nu}^\top {P^\nu}-I_N \right)  \preccurlyeq 0$. Moreover, by Lemma~\ref{lem:NE} ($R=\Sw$),
$
 \| \Pen \|_{\Sw} \le 1 \  
   \Leftrightarrow \ \Pen^\top \Sw\Pen - \Sw \preccurlyeq 0 \  \Leftrightarrow \ \left( {P^\nu} \otimes I_n \right)^\top \left( I_N \otimes S \right) \left( {P^\nu} \otimes I_n \right) -  I_N \otimes S  \preccurlyeq 0
   \Leftrightarrow \ \left( {P^\nu}^\top \otimes I_n^\top \right) \left( I_N \otimes S \right) \left( {P^\nu} \otimes I_n \right) -  I_N \otimes S  \preccurlyeq 0 
   \Leftrightarrow \ \left( {P^\nu}^\top I_N {P^\nu} \otimes I_n^\top S I_n \right) -  I_N \otimes S  \preccurlyeq 0 
    \Leftrightarrow \ \left( {P^\nu}^\top {P^\nu} \right) \otimes  S   -  I_N \otimes S  \preccurlyeq 0 
    \Leftrightarrow \  \left( {P^\nu}^\top {P^\nu}-I_N \right) \otimes S  \preccurlyeq 0.$
Since $S\succ 0$ and $\left( {P^\nu}^\top {P^\nu}-I_N \right)  \preccurlyeq 0$, by the properties of the Kronecker product  we conclude that $\left( {P^\nu}^\top {P^\nu}-I_N \right) \otimes S  \preccurlyeq 0$. Finally, by the previous equivalence, we have $\| \Pen \|_{\Sw} \le 1$.
\end{proof}

Note that a single iteration of Algorithm~\ref{alg:mem} updates the signal $\ze_{(k)}=\left[z^1_{(k)};\ \hdots; z^N_{(k)}\right]$ according to the mapping
\begin{equation}\label{eq:mnt}
 \mnt(\ze):=\Pen_{_2} \xe(\Pen_{_1}\ze)
\end{equation}
and $\mnz\equiv\mn$. The following proposition characterizes the regularity properties of $\mnt$ for different choices of  $\nu_1,\nu_2$. 
\begin{proposition}[Regularity of the update mapping] \label{thm:RegularityConsensusOptimizers} 
 The following statements hold.

\begin{enumerate}
\item If~$M_i \succ 0\ \forall i\in\Z[1,N]$ and $\|P\|\le1$, then the mapping $\mnz\equiv\mn$  in~\eqref{eq:mn} is a CON in $\mathcal{H}_{I_N\otimes Q}$;
\item  If~$M_i \succcurlyeq 0\ \forall i\in\Z[1,N]$ and $\|P\|\le1$, then the mapping $\mnz\equiv\mn$ in~\eqref{eq:mn} is NE in $\mathcal{H}_{I_N\otimes Q}$;

\item If $-q_iQ \preccurlyeq \C \prec 0\ \forall i\in\Z[1,N]$, $\nu\in2\N$ and $P=P^\top$, then the mapping $\mnh$ in~\eqref{eq:mnt}  is FNE in $\mathcal{H}_{I_N\otimes (-C)}$; 
\item If $0 \prec \C$, $\nu\in2\N$ and $P=P^\top$, then the mapping $\mnh$ in~\eqref{eq:mnt} is SPC in $\mathcal{H}_{I_N\otimes C}$. \hspace{2.2cm}$\square$
\end{enumerate}
\end{proposition}
\begin{proof}
\begin{enumerate} 
\item Let $\Sw:=I_N\otimes Q$.
By Lemma~\ref{lemma:RegularityBigOptimizer}, $\xe$ is a CON in $\mathcal{H}_{\bold{S}}$ and, by Lemmas~\ref{lem:NE} and~\ref{lemma:NE_is_preserved}, $\Pen$ is NE in $\mathcal{H}_{\bold{S}}$.
Hence the mapping ${\boldsymbol{\mathcal{A}}_\nu}= \Pen \xe$, composition of a CON mapping and a NE one, is a CON in $\mathcal{H}_{\bold{S}}$.

\item Analogous to the proof of statement 1, with $\xe$ NE.

\item Let  $\Sw:=I_N\otimes (-C)$. From Lemma~\ref{lemma:RegularityBigOptimizer},  $\xe$ is FNE in $\mathcal{H}_{\bold{S}}$. Since $P$ is by construction a row-stochastic matrix and $P=P^\top$, $P$ is a doubly stochastic matrix. Consequently $\|P\|_1=\|P\|_{\infty}= 1$. From H\"{o}lder's inequality, we have $\|P\|\le \sqrt{\|P\|_1\cdot\|P\|_{\infty}} = 1$. Therefore, from Lemma~\ref{lemma:NE_is_preserved}, $\|\Penh\|_{\Sw}\le1$. Moreover, $P=P^\top$ implies $\Penh=\Penh^\top$ and $\Penh\Sw=({P^{\frac{\nu}{2}}} \otimes I_n)(I_N \otimes S)=({P^{\frac{\nu}{2}}}  \otimes S)=\Sw\Penh$. Therefore for any $\ae,\be \in \mathbb{R}^{Nn}$,
\begin{align}\label{eq:fne_a}
&\|\mnh(\ae)-\mnh( \be)\|^2_{\Sw}\\&=\notag\|\Penh(\xe(\Penh (\ae)))-\Penh(\xe(\Penh( \be)))\|^2_{\Sw}\\\notag&\le \|\Penh\|^2_{\Sw}\ \|\xe(\Penh (\ae))-\xe(\Penh( \be))\|^2_{\Sw}\\\notag&\le  \|\xe(\Penh (\ae))-\xe(\Penh( \be))\|^2_{\Sw}\\\notag& \le \left( \Penh \ae-\Penh\be\right)^\top{\Sw}\left( \xe(\Penh (\ae))-\xe(\Penh( \be))\right)\\\notag&=\left( \ae-\be\right)^\top\Penh^\top{\Sw}\left(  \xe(\Penh (\ae))-\xe(\Penh( \be))\right)
\\\notag&=\left( \ae-\be\right)^\top  {\Sw}\Penh\left(  \xe(\Penh (\ae))-\xe(\Penh( \be))\right)
\\\notag&=\left( \ae-\be\right)^\top {\Sw}\left( \mnh(\ae)-\mnh(\be)\right),
\end{align}
where the third inequality derives from $\xe$ being FNE.  The inequality in~\eqref{eq:fne_a} implies that $\mnh (\cdot)$ is FNE in $\mathcal{H}_{\Sw}$, by Lemma~\ref{lem:FNE}. 
\item Let  $\Sw:=I_N\otimes C$. From Lemma ~\ref{lemma:RegularityBigOptimizer}, $-\xe(\cdot)$ is  MON in $\mathcal{H}_{\Sw}$. Moreover $P=P^\top$ implies that $\Penh^\top\Sw=\Sw\Penh$. Therefore for any $\ae,\be \in \mathbb{R}^{Nn}$,
\begin{align*}
&\left( -\mnh (\ae)+\mnh( \be)\right)^\top \Sw \left( \ae-\be\right) \\&=\left( -\Penh(\xe(\Penh (\ae)))+\Penh(\xe(\Penh( \be)))\right)^\top\!\!\Sw\left( \ae-\be\right) \\
&=\left( -\xe(\Penh (\ae))+\xe(\Penh( \be))\right)^\top\Penh^\top\Sw\left( \ae-\be\right)\\
&=\left( -\xe(\Penh (\ae))+\xe(\Penh( \be))\right)^\top\Sw\left(\Penh \ae-\Penh\be\right)\\
&=\left( -\xe(\tilde \ae)+\xe(\tilde \be)\right)^\top\Sw\left(\tilde \ae-\tilde\be\right)\ge 0,
\end{align*}
which implies that $-\mnh (\cdot)$ is MON in $\mathcal{H}_{\Sw}$.  Since $I_{Nn}(\cdot)$ is SMON, it follows  from Lemma~\ref{lem:SA}  that $I_{Nn}(\cdot)-\mnh(\cdot)$ is SMON and  Lipschitz, since it is the composition of Lipschitz mappings. Consequently, $\mnh (\cdot)$ is SPC in $\mathcal{H}_{\bold{S}}$, by Lemma~\ref{lem:SA}.
\end{enumerate}

\end{proof}

We are now ready to prove Theorem~\ref{convergence}. We start by noting that, since the sets $\mathcal{X}^i$ are convex and compact for every $i\in\Z\left[1,N\right]$, the mapping $\m_{\nu_1,\nu_2}:\Pen_{_{2}}\mathcal{X}_{1\times N}\rightarrow \Pen_{_2}\mathcal{X}_{1\times N}$ is defined from a compact and convex set  to itself. For simplicity let  $\boldsymbol{a}_{(k)}\left(\nu_1,\nu_2,\ze_0\right):=\left[ \mathcal{A}^1_{\nu_1,0}(\ze_{(k)}); \, \hdots ;\, \mathcal{A}^N_{\nu_1,0}(\ze_{(k)}) \right]$.

\subsubsection*{Proof of statement 1 and 2 of Theorem~\ref{convergence} }
From Proposition~\ref{thm:RegularityConsensusOptimizers}, under the assumption of statement~1 (or~statement~2), $ \mnz(\cdot)\equiv\mn(\cdot)$ is CON (or NE) in $\mathcal{H}_{I_N\otimes Q}$. Therefore, by using $\bold{\Phi}^{\textup{P--B}}$ (or $\bold{\Phi}^{\textup{K}}$) $\ze_{(k)}$ converges to a fixed point of the mapping $\mn$~\cite[Theorem 2.1]{berinde} (\cite[Theorem 3.2]{berinde}). The proof is immediate upon noticing that, since $\nu_1=0$, $\boldsymbol{a}_{(k)}:=\Pen_{_1}\ze_{(k)}\equiv\ze_{(k)}$. Moreover, a CON mapping has a unique fixed point \cite[Theorem 2.1]{berinde}.
\subsubsection*{Proof of statement 3 and 4 of Theorem~\ref{convergence} } From Proposition~\ref{thm:RegularityConsensusOptimizers}, under the assumption of statement~3 (or~statement~4) the mapping $\ze_{(k)} \mapsto \mnh(\ze_{(k)})$ is FNE (or SPC). Therefore, by using $\bold{\Phi}^{\textup{P--B}}$ (or $\bold{\Phi}_k^{\textup{M}}$) $\ze_{(k)}$ converges to a fixed point $\bar \ze$ of the mapping $\mnh$,
\cite[Section 1, p. 522]{combettes:pennanen:02} (\cite[Fact 4.9, p. 112]{berinde})
. Note that $\boldsymbol{a}_{(k)}:=\Pen_{_1} \ze_{(k)}$ hence, for $\nu_1=\frac{\nu}{2}$, $\boldsymbol{a}_{(k)}$ converges to $\bar{\boldsymbol{a}}:=\Penh \bar \ze$, which is a fixed point of the mapping $\mn$ since
$\bar \ze=\mnh(\bar \ze) \Rightarrow \bar \ze=\Penh \xe(\Penh(\bar \ze)) \Rightarrow \Penh \bar \ze= \Pen \xe(\Penh(\bar \ze)) \Rightarrow \bar{\boldsymbol{a}}=\Pen \xe(\bar{\boldsymbol{a}})\Rightarrow \bar{\boldsymbol{a}}=\mn(\bar{\boldsymbol{a}}).$
\hfill $\blacksquare$

\subsection*{Proof of Corollary~\ref{cor:opinion} (Convergence to a NA Nash equilibrium in the opinion dynamics application)}
For every agent $i\in\Z\left[1,N\right]$, the following equivalence holds
$M_i=\left[\begin{array}{cc}q_iQ & -C \\-C & q_iQ\end{array}\right]
=\left[\begin{array}{cc}(1+\theta_i) & 1 \\1& (1+\theta_i)\end{array}\right]\otimes I_n,\
$
hence the eigenvalues of $M_i$  are $\theta_i$ and $1+\theta_i$, both with multiplicity $n$. It follows that if $\theta_i>0$ then $M_i\succ 0$, if $\theta_i\ge0$ then $M_i\succcurlyeq 0$. 
Consequently, by Theorem~\ref{convergence} the given conditions guarantee convergence of Algorithm~\ref{alg:mem} to a fixed point $\bar \ze$ of the aggregation mapping $\m_1$. Since Assumption~\ref{a3} is satisfied, Theorem~\ref{thm:problem 2} guarantees that  the set of strategies $\left\{x^{i\,\star}(\bar{z}^i)\right\}_{i=1}^N$ is a NA Nash equilibrium.\hfill  $\blacksquare$

\subsection*{Proof of Corollary~\ref{cor:demand_response} (Convergence to an $\varepsilon$-Nash equilibrium in demand-response methods)}
Given the cost function in~\eqref{eq:demand_response} with $p(\bar\sigma_t):=\lambda\bar\sigma_t+p_0$, the following equivalence holds
$
M_i=\left[\begin{array}{cc}q_iQ & -C \\-C & q_iQ\end{array}\right]=
\left[\begin{array}{cc}\rho_i& -\frac{\lambda}{2} \\-\frac{\lambda}{2}& \rho_i\end{array}\right]\otimes I_T,$
hence the eigenvalues of $M_i$ are $\rho_i+\frac{\lambda}{2}$ and $\rho_i-\frac{\lambda}{2}$, both with multiplicity $T$. It follows that if $\rho_i-\frac{\lambda}{2}>0$, then $M_i\succ 0$ while if $\rho_i-\frac{\lambda}{2}\ge0$  then $M_i\succcurlyeq 0$. Moreover, $C=\frac{\lambda}{2} I_T\succ 0$.
Consequently, by Theorem~\ref{convergence} the given conditions guarantee convergence of Algorithm~\ref{alg:mem} to a fixed point $\bar \ze$ of the aggregation mapping $\mn$. The conclusion follows from  Theorem~\ref{thm:problem 1}. \hfill $\blacksquare$

\subsection*{Hierarchical control}
\begin{lemma}
If $P_M\in\R^{M\times M}$ satisfies Assumption~\ref{a2}, then also $P=P_M\otimes \frac1B \mathbbm{1}_{B}\mathbbm{1}_{B}^\top$ does. Moreover, if $\|P_M\|\le1$, then $\|P\|\le1.$
\label{lem:matrix}
\end{lemma}
\begin{proof}
By the properties of the Kronecher product we have
$\textstyle\lim_{\nu\rightarrow\infty}\left(P_M\otimes \frac1B \mathbbm{1}_{B}\mathbbm{1}_{B}^\top\right)^\nu  = \textstyle\lim_{\nu\rightarrow\infty}\left(P_M^\nu\otimes \frac1B \mathbbm{1}_{B}\mathbbm{1}_{B}^\top\right) = \textstyle\left(\frac1M \mathbbm{1}_{M}\mathbbm{1}_{M}^\top\right)\otimes\left(\frac1B \mathbbm{1}_{B}\mathbbm{1}_{B}^\top\right)=\textstyle\frac{1}{MB} \mathbbm{1}_{MB}\mathbbm{1}_{MB}^\top=\ones.$ Moreover,
$
\|P\|=\sqrt{\rho(P^\top P)}= 
 \textstyle\sqrt{\rho\left(\left(P_M^\top\otimes \frac1B \mathbbm{1}_{B}\mathbbm{1}_{B}^\top\right)\left(P_M\otimes \frac1B \mathbbm{1}_{B}\mathbbm{1}_{B}^\top\right)\right)}
= \textstyle\sqrt{\rho\left(P_M^\top P_M\otimes \frac1B \mathbbm{1}_{B}\mathbbm{1}_{B}^\top\right)}=\sqrt{\rho\left(P_M^\top P_M\right)}=\|P_M\|\le1,$
where we used the fact that the matrix $\frac1B \mathbbm{1}_{B}\mathbbm{1}_{B}^\top$ has all the eigenvalues in $0$  except one which is equal to $1$, all the eigenvalues of $P_M^\top P_M$ are nonnegative real and the eigenvalues of the Kronecker product are the product of the eigenvalues of the two matrices.
\end{proof}
\section*{Appendix C}
From the proof of Theorem~\ref{thm:problem 1} for any $\varepsilon>0$ and $N>\bar N:=\frac{K}{\varepsilon}$, the set  of strategies $\{x^{i\,\star}(\bar{z}^i)\}_{i=1}^N$ is  MF $\varepsilon$-Nash equilibrium if $\| P^{ \nu}-\ones\|_\infty   \le {\varepsilon_d}:=\frac{\varepsilon}{K}-\frac1N$.
\begin{algorithm} 
\caption{Precomputing $\bar \nu$ on a given network $P$}
\label{alg:nu}
\textbf{Initialization}: $k \leftarrow 0$, fix $\varepsilon_d>0$, $s^i\leftarrow 0\ \forall i, P^0=I_n$.
\textbf{Iterate until $\boldsymbol{s^i=1}$ }: Each agent $i$ synchronously
\begin{enumerate}
\item receives the weight coefficients $\{P^k_{hj}\}_{j=1}^N$ from each\\ of its neighbors $h\in\mathcal{N}^i$
\item computes $\{P^{k+1}_{ij}=\sum_{h\in\mathcal{N}^i} P_{ih}P^k_{hj} \}_{j=1}^N$
\item if $\sum_{j=1}^N | P_{ij}^{k+1}-\frac{1}{N}|<\varepsilon_d $, then $s^i\leftarrow 1$, else  $s^i \leftarrow 0$
\item repeats $N$ times
$s^i \leftarrow \min_{j\in\mathcal{N}^i\cup\{i\}}\{{s^j}\}$
\end{enumerate}
$k \leftarrow k+1$\\
\textbf{Output}:  $\bar \nu \leftarrow k$.
\end{algorithm}
\vspace{0.2cm}
If the network is strongly connected and Assumption~4 holds, then it is possible to compute, for any $\varepsilon_d>0$ and in a distributed fashion, a value of $\bar \nu$ such that $\| P^{\bar \nu}-\ones\|_\infty   \le {\varepsilon_d}$ by using Algorithm~\ref{alg:nu}.
The  proof is based on the following fact: suppose that each agent $i$ has a value of $s^i\in\{0,1\}$. Since the graph is strongly connected after at most $N$ iterations of the update 
$s^i \leftarrow \min_{j\in\mathcal{N}^i\cup\{i\}}\{{s^j}\}$ the agents reach consensus to a value $s$ which is $1$ if all the $s^i$ at the beginning were equal to $1$ and is $0$ otherwise. Therefore if Algorithm~\ref{alg:nu} terminates it must be at a value $\bar \nu$ for which  $\sum_{j=1}^N | P_{ij}^{\bar \nu}-\frac{1}{N}|<\varepsilon_d $ for all $i$, which in turn implies $\| P^{\bar \nu}-\ones\|_\infty   \le {\varepsilon_d}$. Note that Assumption~4 guarantees that for each agent $i$, there exists a value $\bar\nu^i$ such that $\sum_{j=1}^N | P_{ij}^k-\frac{1}{N}|<\varepsilon_d$ for all $k\ge \bar\nu^i$. Hence for $k>\max\{\bar{\nu}^i\}_{i=1}^N$, $s^i=1$ for all $i$ and  Algorithm~\ref{alg:nu} is guaranteed to terminate in a finite number of iterations.


\end{document}